# TTRG integration of light transport equations:

# Azymuthally integrated radiances inside a Lambertian foliage

Sophie Royer
*Institut für Botanik, Universität Innsbruck, Austria*
Antoine Royer
*Département de Génie Physique, Ecole Polytechnique, Montréal, Québec H3C 3A7, Canada*

A method for numerically integrating transport equations, combining transfer matrices, transmission-reflection matrices, and Green's matrices (TTRG), was recently proposed. The present paper deals specifically with azymuthally integrated radiances inside a horizontally homogeneous canopy of Lambertian leaves. Its main purpose is to test the accuracy of TTRG by applying it to non-trivial models possessing analytical solutions, that were given in another paper. Comparison is made with the widely used iterative integration (or 'relaxation' method). Cases of extreme light trapping are given for which iterative integration is hardly practical, while TTRG remains as accurate and rapid.

## 1. Introduction

In a preceding paper [1], a method was proposed for numerically integrating transport equations by combining transfer matrices, transmission-reflection matrices, and Green's matrices (TTRG). The purpose of the present paper is to test the accuracy and speed of TTRG, by applying it to analytically solvable models that were formulated in another paper [2], and compare with the widely used iterative integration (or 'relaxation' method [3]).

The physical system considered is light propagating in a horizontally homogeneous, azymuthally symmetric ($\varphi$-symmetric for short) canopy of Lambertian leaves [3-5]. In the present paper, we only compute $\varphi$-integrated radiances (i.e., radiances integrated over the azymuthal angle $\varphi$), which is much easier. These give the complete solution if the incident radiance is $\varphi$-symmetric (as skylight is usually assumed to be), since the radiance inside the canopy will then also be $\varphi$–symmetric. If there is also sharply directional sunlight incident on the canopy, then the *diffuse* radiance may be assumed, as an approximation [3-5], to be $\varphi$–symmetric, in which case the $\varphi$–integrated radiance again provides the complete solution. Computation of $\varphi$-dependent diffuse radiances is defered to subsequent papers.

Our treatmen is more general than usual in that we allow the top and under sides of leaves to have different optical properties, which is more realistic. Moreover, this allows us to create artificial situations, such as extreme light trapping, for testing the limits of numerical methods.

We assume purely Lambertian scattering and emission of light, both by individual leaves and by the ground surface. The light transport equation (LTE) can then be discretized over photon directions in an *analytic* way. This makes high precision tests possible.

High precision tests are done in cases with zero emission (either thermal emission, or first-scattered-sunlight 'emission'), by applying TTRG to non-trivial models possessing analytical solutions [2]. Agreement is generally better than 10 significant digits when we let the computer carry 15 digits (double precision). With emission present, one must numerically integrate 'propagated emissions'[1], so that the overall precision of TTRG is limited by the accuracy of that numerical integration.

Cases of extreme light trapping are given for which interative integration requires on the order of hours to attain accuracies of $10^{-3}$ or so, while TTRG acheives $10^{-10}$ and better in fractions of a second. Also, just stipulating a convergence criterium for the iterative integration can be problematic. Thus, there are situations where iterative integration is hardly practical, while TTRG remains as efficient. Such problems do not occur with *realistic* canopies, but TTRG is still appreciably faster, as well as much more accurate.

The paper has three parts: Part I deals with the more formal aspects of the paper, including the derivation of the $\varphi$-integrated LTE, and its discretization over photon inclinations. Part II roughly estimates the appropriate thicknesses for the 'medium' layers required in TTRG, and for the 'thin' layers used in both TTRG and iterative integration. Part III presents and discusses the results of numerical computations. Appendices contain technical details.

*Remark*: Equations in references [1] and [2] will be refered to as Eq.I-(10), or Eq.II-(20), etc. Also, section 3 of Ref.[1] will be refered to as section I-3, etc.

***Notation and conventions***: We let the positive $z$ direction point *downwards*. A hat over a scalar $c$ signifies absolute value, $\hat{c} \equiv |c| > 0$. Using the step function $\Theta(x)$, we denote

$$|x|_{\pm} \equiv |x|\Theta(\pm x) = \hat{x}\Theta(\pm x), \qquad \Theta(x>0)=1, \quad \Theta(x<0)=0 \tag{1}$$

Unit vectors in direction $\Omega = (\mu, \varphi)$, where $\mu \equiv \cos\theta$, are denoted $\mathbf{\Omega}$:

$$\begin{aligned} &\mathbf{\Omega} = \left(\Omega_x, \Omega_y, \Omega_z\right) = \left(\sin\theta\cos\varphi, \sin\theta\sin\varphi, \mu\right), \qquad \mu = \Omega_z = \cos\theta \\ &d\Omega = \sin\theta d\theta d\varphi = d\mu d\varphi, \qquad \int d\Omega\, f(\Omega) = \int_{-\pi}^{\pi} d\varphi \int_{-1}^{1} d\mu\, f(\mu,\varphi) \end{aligned} \tag{2}$$

The scalar product of two unit vectors $\mathbf{\Omega}$ and $\mathbf{\Omega}'$ is

$$\mathbf{\Omega}\cdot\mathbf{\Omega}' = \mu\mu' + \sin\theta\sin\theta'\cos(\varphi - \varphi') \tag{3}$$

Dirac $\delta$ functions of solid angles are written

$$\delta(\Omega - \Omega') = \delta(\mu - \mu')\delta(\varphi - \varphi') \tag{4}$$

For any function $f(\mu)$, and any interval $\Delta\mu = (\mu_1, \mu_2)$, we denote

$$f(\Delta\mu) \equiv \int_{\Delta\mu} d\mu\, f(\mu) \equiv \int_{\mu_1}^{\mu_2} d\mu\, f(\mu), \qquad \Delta\mu = (\mu_1, \mu_2), \quad \mu_1 < \mu_2 \tag{5}$$

We note the following expansion for small $\Delta\hat{\mu}$:

$$f(\Delta\mu) = f(\overline{\mu})\Delta\hat{\mu} + \tfrac{1}{24} f''(\overline{\mu})\Delta\hat{\mu}^3 + ..., \qquad \overline{\mu} \equiv \tfrac{1}{2}(\mu_1 + \mu_2), \qquad \Delta\hat{\mu} \equiv \mu_2 - \mu_1 \tag{6}$$

obtained by writing $f(\mu) = f(\overline{\mu}) + f'(\overline{\mu})(\mu - \overline{\mu}) + \frac{1}{2} f''(\overline{\mu})(\mu - \overline{\mu})^2 + ...$ , and noting that

$$\int_{\mu_1}^{\mu_2} d\mu (\mu - \overline{\mu})^n = \int_{-\frac{1}{2}\Delta\hat{\mu}}^{\frac{1}{2}\Delta\hat{\mu}} dx\, x^n = \begin{cases} \frac{2}{n+1}\left(\frac{1}{2}\Delta\hat{\mu}\right)^{n+1} & \text{if } n \text{ is odd} \\ 0 & \text{if } n \text{ is even} \end{cases} \tag{7}$$

We will need the following integrals of the scalar product (3), using the notation (1):

$$\int d\mathbf{\Omega} \left|\mathbf{\Omega}\cdot\mathbf{\Omega}'\right|_{\pm} = \int d\mathbf{\Omega} \left|\mu\right|_{\pm} = \int_{-\pi}^{\pi} d\varphi \int_0^1 \mu d\mu = \pi, \qquad \int d\Omega \left|\mathbf{\Omega}\cdot\mathbf{\Omega}'\right| = 2\pi \tag{8}$$

The following functions are obtained analytically in Appendix B:

$$\begin{aligned} &g_{\pm}(\mu,\mu') = \tfrac{1}{2\pi}\int_{-\pi}^{\pi} d\varphi\, |\mathbf{\Omega}\cdot\mathbf{\Omega}'|_{\pm}, \qquad g_{\pm}(\Delta\mu,\mu') = \int_{\Delta\mu} d\mu\, g_{\pm}(\mu,\mu') \\ &g(\mu,\mu') = \tfrac{1}{2\pi}\int_{-\pi}^{\pi} d\varphi |\mathbf{\Omega}\cdot\mathbf{\Omega}'| = g_+(\mu,\mu') + g_-(\mu,\mu') \end{aligned} \tag{9}$$

We write $\Omega \in D$ or $\mu \in D$ if $\mu > 0$ ($\Omega$ points downwards), $\Omega \in U$ or $\mu \in U$ if $\mu < 0$ (upwards). A function $f(\Omega) = f(\mu)$ independent of $\varphi$ is said to be $\varphi$-symmetric.

## Part I Theoretical aspects

This part presents the more formal aspects of the paper. Section 2 summarizes the results of Ref.[2] that we will need. Section 3 derives the $\varphi$-integrated light transport equation (LTE). Section 4 obtains a $\varphi$-integrated LTE for the *diffuse* light, with first-scattered-sunlight treated as an 'emission'. Section 5 discretizes photon inclinations. Section 6 writes the LTE in matrix form. Section 7 assumes purely Lambertian scattering and emission. Finally, section 8 deals with totally absorbing leaves (black leaves).

## 2 Summary of previous results

We here summarise, for easy reference, results from Ref. [2] that we will use. A photon travelling in direction $\Omega$ is refered to as a 'photon $\Omega$'.

***Leaves***: Scattering and emission coefficients for a *horizontal* leaf, $\sigma(\Omega_i \to \Omega_f)$ and $\varepsilon(\Omega_f)$, are defined such that $\sigma(\Omega_i \to \Omega_f)d\Omega_f$ is the probability for a photon $\Omega_i$ hitting the leaf to get scattered into $d\Omega_f$, and $\varepsilon(\Omega_f)d\Omega_f$ is the number of photons emitted into $d\Omega_f$ by the leaf, per second per unit area of leaf. The absorption coefficient is

$$\alpha(\Omega_i) = 1 - \int d\Omega_f \sigma(\Omega_i \to \Omega_f) \tag{10}$$

It will be convenient to also use the notations

$$s(\Omega_i \to \Omega_f) \equiv \hat{\mu}_i \sigma(\Omega_i \to \Omega_f), \qquad a(\Omega_i) \equiv \hat{\mu}_i \alpha(\Omega_i) \tag{11}$$

Coefficients for the ground surface are written with a subscript $g$, e.g., $s_g(\Omega_i \to \Omega_f)$.

All the above coefficients are assumed invariant under rotations about the vertical $z$ axis. Thus, it suffices to specify the orientation of an inclined leaf by the normal $\Omega_L$ to its underside. For a leaf of orientation $\Omega_L$ (or a 'leaf $\Omega_L$' for short), the optical coefficients are written with a subscript $\Omega_L$, e.g., $s_{\Omega_L}(\Omega_i \to \Omega_f)$.

***Canopy***: A horizontally homogeneous canopy is characterized by volume densities $\eta(\Omega_L, z)d\Omega_L$ of leaf areas oriented within $d\Omega_L$. When a photon $\Omega_i$ travels an infinitesimal distance $\mathbf{\Omega}_i d\ell$ through the canopy (where $d\ell$ is an infinitesimal length), it has probability $\Gamma(\Omega_i, z)d\ell$ to get intercepted by a leaf, and $S(\Omega_i \to \Omega_f, z)d\Omega_f d\ell$ to get scattered into $d\Omega_f$, where

$$\Gamma(\Omega_i, z) = \int d\Omega_L \eta(\Omega_L, z) \left| \mathbf{\Omega}_i \cdot \mathbf{\Omega}_L \right| \tag{12}$$

$$S(\Omega_i \to \Omega_f, z) = \int d\Omega_L \eta(\Omega_L, z)\, s_{\Omega_L}(\Omega_i \to \Omega_f) \tag{13}$$

The probability per unit distance for the photon to get absorbed is, by (10)-(11):

$$A(\Omega_i, z) = \Gamma(\Omega_i, z) - \int d\Omega_f S(\Omega_i \to \Omega_f, z) = \int d\Omega_L\, \eta(\Omega_L, z) a_{\Omega_L}(\Omega_i) \tag{14}$$

The canopy emits into the solid angle $d\Omega_f$, per second per unit volume, a number of photons equal to $\mathcal{E}(\Omega_f, z) d\Omega$, where

$$\mathcal{E}(\Omega_f, z) = \int d\Omega_L \eta(\Omega_L, z) \varepsilon_{\Omega_L}(\Omega_f, z) \tag{15}$$

Here we let the leaf thermal emission $\varepsilon_{\Omega_L}(\Omega_f, z)$ depend on $z$, because the temperature of a leaf will depend in general on its position inside the canopy.

***Radiances***: We denote by $I(\Omega, z) d\Omega$ the photon flux in direction $\Omega$, i.e., the number of photons within $d\Omega$ crossing per second a unit area perpendicular to $\Omega$. We will often call $I(\Omega, z)$ the 'radiance' (though more properly radiance is the energy flux). Radiances in downwards ($\Omega \in D$) and in upwards ($\Omega \in U$) directions will also be denoted by $D(\Omega, z)$ and $U(\Omega, z)$, so that

$$I(\Omega, z) = D(\Omega, z) \quad \text{if} \quad \Omega \in D, \qquad I(\Omega, z) = U(\Omega, z) \quad \text{if} \quad \Omega \in U \tag{16}$$

***The LTE***: Over an infinitesimal distance $\mathbf{\Omega}_f d\ell$, the flux $I(\Omega_f, z) d\Omega_f$ looses photons due to interception by leaves, but gains photons, either emitted into $d\Omega_f$, or scattered into $d\Omega_f$ from other directions $\Omega_i$. Whence the light transport equation (since $d\ell = dz/\mu_f$):

$$\begin{aligned} &\mu_f \frac{d}{dz} I(\Omega_f, z) = -\Gamma(\Omega_f, z) I(\Omega_f, z) + \int d\Omega_i\, I(\Omega_i, z) S(\Omega_i \to \Omega_f, z) + \mathcal{E}(\Omega_f, z) \\ &\hat{\mu}_f U(\Omega_f, z_g) = \int_{\mu_i \in D} d\Omega_i\, D(\Omega_i, z_g)\, s_g(\Omega_i \to \Omega_f) + \varepsilon_g(\Omega_f), \qquad \mu_f < 0 \end{aligned} \tag{17}$$

The second line is the reflection and emission by the ground surface, at $z = z_g$.

### 3 φ-symmetric canopy, $\varphi$-integrated LTE

We henceforth assume $\varphi$-symmetric leaf area densities. The *$\varphi$-integrated* radiance then satisfies an equation free of azymuthal angles, much simpler than the full LTE (17).

**φ-symmetric canopy**: So let us assume $\varphi$-symmetric leaf area densities

$$\eta(\Omega_L,z)=\tfrac{1}{2\pi}\tilde{\eta}(\mu_L,z) \qquad \text{where} \qquad \tilde{\eta}(\mu_L,z)\equiv\int_{-\pi}^{\pi}d\varphi_L\eta(\Omega_L,z) \tag{18}$$

Since these define no prefered azymuthal direction, $\Gamma(\Omega,z)$ and $S(\Omega_i\to\Omega_f,z)$ will also be $\varphi$-symmetric, that is:

$$\begin{aligned}&\Gamma(\Omega,z)=\Gamma(\mu,z), &&S\big(\Omega_i\to\Omega_f,z\big)=S(\mu_i\to\mu_f,\varphi_{if},z)\\ &s_g(\Omega_i\to\Omega_f)=s_g(\mu_i\to\mu_f,\varphi_{if}), &&\varphi_{if}\equiv\varphi_i-\varphi_f\end{aligned} \tag{19}$$

where we also assume $\varphi$-symmetric ground scattering coefficients $s_g$. Using (9), we have

$$\Gamma(\mu,z)=\int_{-1}^{1}d\mu_L\tilde{\eta}(\mu_L,z)g(\mu,\mu_L) \tag{20}$$

***φ-integrated LTE***: If the light incident on the canopy is $\varphi$–symmetric, then so will the radiance inside the canopy, since there is then no prefered azymuthal direction whatsoever, i.e.:

$$I(\Omega,z)=\tfrac{1}{2\pi}\tilde{I}(\mu,z) \qquad \text{where} \qquad \tilde{I}(\mu,z)\equiv\int_{-\pi}^{\pi}d\varphi\, I\big(\Omega,z\big) \tag{21}$$

The $\varphi$–integrated radiance $\tilde{I}(\mu,z)$ is much easier to compute than $I(\Omega,z)$, for it satisfies a much simpler transport equation, free of azymuthal angles. Indeed, integrating (17) over $\varphi_f$, and using (19), we obtain:[1]

$$\begin{aligned}&\mu_f\frac{d}{dz}\tilde{I}(\mu_f,z)=-\Gamma(\mu_f,z)\tilde{I}(\mu_f,z)+\int_{-1}^{1}d\mu_i\tilde{I}(\mu_i,z)\tilde{S}(\mu_i\to\mu_f,z)+\tilde{\mathcal{E}}(\mu_f,z)\\ &\hat{\mu}_f\tilde{U}(\mu_f,z_g)=\int_{\mu_i\in D}d\mu_i\,\tilde{D}(\mu_i,z_g)\tilde{s}_g(\mu_i\to\mu_f)+\tilde{\varepsilon}_g(\mu_f) \qquad (\mu_f\in U)\end{aligned} \tag{22}$$

where a tilde signifies integration over $\varphi$. In particular:

$$\begin{aligned}&\tilde{S}(\mu_i\to\mu_f,z)\equiv\int_{-\pi}^{\pi}d\varphi S(\mu_i\to\mu_f,\varphi,z)=\int_{-1}^{1}d\mu_L\tilde{\eta}(\mu_L,z)\tilde{\tilde{s}}_{\mu_L}(\mu_i\to\mu_f)\\ &\tilde{\mathcal{E}}(\mu,z)\equiv\int_{-\pi}^{\pi}d\varphi\mathcal{E}(\Omega,z)=\int_{-1}^{1}d\mu_L\tilde{\eta}(\mu_L,z)\tilde{\tilde{\varepsilon}}_{\mu_L}(\mu,z)\end{aligned} \tag{23}$$

---

[1] Note that $\tilde{I}(\mu,z)$ satisfies a *closed* equation only if the canopy is *globally* $\varphi$–symmetric. If only *local* $\varphi$–symmetry (19) holds, without horizontal homogeneity, then the integral over $\varphi_f$ of $\mathbf{\Omega}_f\cdot\nabla I(\mathbf{\Omega}_f,\mathbf{r})$, in Eq.(22) of Ref.[2], does not produce $\tilde{I}(\mu_f,\mathbf{r})$. Indeed, we do not expect $\varphi$–symmetric radiances inside a canopy of parallel hedges, even if (19) holds locally.

where we define $\varphi_f$-integrated, $\varphi_L$-averaged leaf coefficients

$$\begin{aligned}\tilde{\bar{s}}_{\mu_L}(\mu_i \to \mu_f) &\equiv \tfrac{1}{2\pi}\int_{-\pi}^{\pi} d\varphi_L \int_{-\pi}^{\pi} d\varphi_f \, s_{\boldsymbol{\Omega}_L}(\Omega_i \to \Omega_f) \\ \tilde{\bar{\varepsilon}}_{\mu_L}(\mu, z) &\equiv \tfrac{1}{2\pi}\int_{-\pi}^{\pi} d\varphi_L \int_{-\pi}^{\pi} d\varphi_f \, \varepsilon_{\boldsymbol{\Omega}_L}(\Omega, z)\end{aligned} \tag{24}$$

***Canopy absorption and emission***: The local canopy absorption coefficients (14) may be written, using (10)-(11):

$$A(\mu, z) = \Gamma(\mu, z) - \int d\mu_f \tilde{S}(\mu_i \to \mu_f, z) = \int d\mu_L \tilde{\eta}(\mu_L, z)\bar{a}_{\mu_L}(\mu_i) \tag{25}$$

where we define $\varphi_L$-averaged leaf absorption coefficients, using (9):

$$\bar{a}_{\mu_L}(\mu_i) \equiv \tfrac{1}{2\pi}\int_{-\pi}^{\pi} d\varphi_L a_{\Omega_L}(\Omega_i) = g(\mu_i, \mu_L) - \int d\mu_f \tilde{\bar{s}}_{\mu_L}(\mu_i \to \mu_f) \tag{26}$$

Here we noted that $a_{\Omega_L}(\Omega_i)$ can only be a function of the angle between $\Omega_i$ and $\Omega_L$, hence of $\varphi_i - \varphi_L$, by (3). Note also that the ground absorption coefficient is given by

$$a_g(\Omega_i) = a_g(\mu_i) = \hat{\mu}_i - \int_{\mu_f \in U} d\mu_f \tilde{s}_g(\mu_i \to \mu_f) \tag{27}$$

The total numbers of photons absorbed or emitted per second, per unit volume of canopy, or per unit area of ground surface, are:

$$\begin{aligned}&\mathcal{A}(z) = \int d\Omega \, I(\Omega, z) A(\mu, z) = \int d\mu \, \tilde{I}(\mu, z) A(\mu, z), \qquad \mathcal{E}(z) = \int d\Omega \, \mathcal{E}(\Omega, z) \\ &\mathcal{A}_g = \int_{\mu \in D} d\mu \, \tilde{D}(\mu, z_g) a_g(\mu), \qquad \mathcal{E}_g \equiv \int_{\mu \in U} d\mu \, \tilde{\varepsilon}_g(\mu)\end{aligned} \tag{28}$$

***Continuity equation***: Integrating (22) over $\mu_f$, using (25), yields continuity equations

$$\begin{aligned}&\frac{d}{dz} i(z) = \mathcal{E}(z) - \mathcal{A}(z), \qquad i(z) \equiv \mathcal{D}(z) - \mathcal{U}(z) \\ &i_g \equiv \mathcal{D}_g - \mathcal{U}_g = \mathcal{E}_g - \mathcal{A}_g\end{aligned} \tag{29}$$

where $i(z)$ is the *net* vertical flux, and $\mathcal{D}(z)$ and $\mathcal{U}(z)$ are 'down' and 'up' *vertical* fluxes (numbers of photons crossing per second a horizontal unit area):

$$\mathcal{D}(z) \equiv \int_{\mu \in D} d\mu \, \hat{\mu} \, \tilde{D}(\mu, z), \qquad \mathcal{U}(z) \equiv \int_{\mu \in U} d\mu \, \hat{\mu} \, \tilde{U}(\mu, z) \tag{30}$$

In (29) we used the shorthands $i_g = i(z_g)$, $\mathcal{D}_g = \mathcal{D}(z_g)$, etc.

## 4 First-scattered sunlight treated as an 'emission'

In this section we consider the effect of sunlight, and how to deal with it. Light incident on the top $z = z_0$ of the canopy comprises skylight and sunlight. Diffuse skylight is assumed isotropic in 'down' directions, while sunlight is modeled as a sharply directional radiance in the (downwards) direction $\Omega_h$ of sunrays (here $h$ stands for *helios*, and $d$ for *diffuse*):

$$D^d(\Omega, z_0) = D_0^d, \qquad D^h(\Omega, z_0) = h_0 \delta(\Omega - \Omega_h) \tag{31}$$

The incident *vertical* fluxes are thus, by (30) and (4) (writing $\mathcal{D}_0 \equiv \mathcal{D}(z_0)$, etc.):

$$\mathcal{D}_0^d = \int_{\mu \in D} d\mu \, \hat{\mu} \, \tilde{D}^d(\mu, z_0) = \pi D_0^d, \qquad \mathcal{D}_0^h = \hat{\mu}_h h_0 \tag{32}$$

Inside the canopy, *direct* sunlight $I^h(\Omega, z)$ is attenuated due to interception by leaves, and is given by:

$$\begin{aligned} &I^h(\Omega, z) = D^h(\Omega, z) = h(z)\delta(\Omega - \Omega_h), \qquad h(z) \equiv h_0 \xi_{\mu_h}(z) \\ &\xi_\mu(z) \equiv \xi_\mu(z_0, z) \equiv e^{-\Lambda_\mu(z_0, z)}, \qquad \Lambda_\mu(z_1, z_2) \equiv \int_{z_1}^{z_2} dz \, \Gamma(\mu, z)/\mu \end{aligned} \tag{33}$$

***Broken* φ*-symmetry*** : In general, due the sharp directionality of sunlight, the radiance $I(\Omega, z)$ will depend on $\varphi$, and so will the thermal emission $\mathcal{E}(\Omega, z)$ (since the temperature of a leaf depends on its orientation, especially relative to sunlight). Of course, even if $I(\Omega, z)$ is not $\varphi$-symmetric, the $\varphi$-integrated LTE (22) is still valid. Its input is the $\varphi$-integrated incident radiance

$$\tilde{D}(\mu, z_0) = \tilde{D}^d(\mu, z_0) + \tilde{D}^h(\mu, z_0) = 2\mathcal{D}_0^d + h_0 \delta(\mu - \mu_h) \tag{34}$$

by (31)-(32) and (4). But $\tilde{I}(\mu, z)$ no longer yields $I(\Omega, z)$ via (21).

***Diffuse light***: However, the *diffuse* radiance

$$I^d(\Omega, z) \equiv I(\Omega, z) - I^h(\Omega, z) = I(\Omega, z) - h(z)\delta(\Omega - \Omega_h) \tag{35}$$

is expected to vary mildly with $\varphi$, and is often approximated as $\varphi$-symmetric [3-5]:

$$\begin{aligned} &I^d(\Omega, z) \approx \tfrac{1}{2\pi} \tilde{I}^d(\mu, z) = \tfrac{1}{2\pi} \tilde{I}(\mu, z) - \tfrac{1}{2\pi} \tilde{I}^h(\mu, z) \\ &\tilde{I}^h(\mu, z) = h(z)\delta(\mu - \mu_h) \end{aligned} \tag{36}$$

Within that approximation, $\tilde{I}(\mu, z)$ still provides the complete solution, since

$$I(\Omega, z) = \tfrac{1}{2\pi} \tilde{I}^d(\mu, z) + I^h(\Omega, z) = \tfrac{1}{2\pi} \tilde{I}(\mu, z) - \tfrac{1}{2\pi} \tilde{I}^h(\mu, z) + I^h(\Omega, z) \tag{37}$$

The vertical fluxes are

$$\mathcal{D}(z) = \int_{\mu\in D} d\mu\,\hat{\mu}\,\tilde{D}^d(\mu,z) + \hat{\mu}_h h(z), \qquad \mathcal{U}(z) = \int_{\mu\in U} d\mu\,\hat{\mu}\,\tilde{U}^d(\mu,z) \tag{38}$$

The total absorption rates in (28) are now given by

$$\begin{aligned} \mathcal{A}(z) &= \int d\mu\,\tilde{I}^d(\mu,z)A(\mu,z) + h(z)A(\mu_h,z) \\ \mathcal{A}_g &= \int_{\mu_i\in D} d\mu_i\,\tilde{D}^d(\mu_i,z_g)a_g(\mu_i) + h(z_g)a_g(\mu_h) \end{aligned} \tag{39}$$

***φ-integrated diffuse LTE***: In practice, photon inclinations $\mu$ must be discretized, to enable numerical computation. Now, the attenuation (33) of sunlight, and the vertical incident sunlight flux $\mu_h h_0$ (hence the energy flow into the canopy), are very sensitive to the inclination $\mu_h$ of sun rays. So tinkering with the latter (e.g., putting it equal to one of the discrete inclinations) may entail sizable errors, because sunlight is so intense. It is therefore customary [3-5] to compute accurately the attenuated sunlight $h(z)$, thence the sunlight that has been scattered *once*, and treat that as an 'emission'. This is achieved by substituting $\tilde{I}(\mu,z) = \tilde{I}^d(\mu,z) + \tilde{I}^h(\mu,z)$ into (22) (or by integrating the diffuse LTE (50) in Ref. [2] over $\varphi_f$), to obtain

$$\begin{aligned} \mu_f \frac{d}{dz}\tilde{I}^d(\mu_f,z) &= -\Gamma(\mu_f,z)\tilde{I}^d(\mu_f,z) + \int_{-1}^{1} d\mu_i \tilde{I}^d(\mu_i,z)\tilde{S}(\mu_i \to \mu_f,z) + \tilde{\mathcal{E}}^{tot}(\mu_f,z) \\ \hat{\mu}_f \tilde{U}^d(\mu_f,z_g) &= \int_{\mu_i\in D} d\mu_i\,\tilde{D}^d(\mu_i,z_g)\tilde{s}_g(\mu_i \to \mu_f) + \tilde{\varepsilon}_g^{tot}(\mu_f), \qquad \mu_f < 0 \end{aligned} \tag{40}$$

where $\tilde{\mathcal{E}}^{tot} = \tilde{\mathcal{E}} + \tilde{\mathcal{E}}^h$ and $\tilde{\varepsilon}_g^{tot} = \tilde{\varepsilon}_g + \tilde{\varepsilon}_g^h$ comprise 'true' (i.e., thermal) emissions $\tilde{\mathcal{E}}$ and $\tilde{\varepsilon}_g$, and first-scattered-sunlight 'emissions' given by, in view of (23)-(24):

$$\begin{aligned} \tilde{\mathcal{E}}^h(\mu_f,z) &= h(z)\tilde{S}(\mu_h \to \mu_f,z) = \int_{-1}^{1} d\mu_L \tilde{\eta}(\mu_L,z)\tilde{\tilde{\varepsilon}}_{\mu_L}^h(\mu_f,z) \\ \tilde{\tilde{\varepsilon}}_{\mu_L}^h(\mu_f,z) &= h(z)\tilde{\tilde{s}}_{\mu_L}(\mu_h \to \mu_f), \qquad \tilde{\varepsilon}_g^h(\mu_f) = h(z_g)\tilde{s}_g(\mu_h \to \mu_f) \end{aligned} \tag{41}$$

## 5 Discrete photon inclinations

We now discretize the photon inclinations $\mu \in (-1,1)$. This must be done carefully, in order to preserve energy balance. We use two different methods. Which of the two is used matters little in practice, but does matter in high precision tests.

***Photon inclination sectors***: Partition the interval $(-1,1)$ into $N_J = N_D + N_U$ sectors $\Delta\mu_j$:

$$\begin{aligned} &(-1,1) = (\mu_0,\mu_1,\ldots,\mu_{N_J}), \qquad \mu_0 = -1, \quad \mu_{N_D} = 0, \quad \mu_{N_J} = 1 \\ &\Delta\mu_j = (\mu_{j-1},\mu_j), \qquad \bar{\mu}_j = \tfrac{1}{2}(\mu_{j-1} + \mu_j), \qquad j = 1,2,\ldots,N_J \end{aligned} \tag{42}$$

where $\overline{\mu}_j$ are mean inclinations. Note that no sector straddles $\mu = 0$. We write $j \in D$ if $\overline{\mu}_j > 0$ is downwards, and $j \in U$ if $\overline{\mu}_j < 0$ is upwards. Denoting

$$\Delta\hat{\mu}_j \equiv \left|\mu_j - \mu_{j-1}\right|, \qquad \Delta\mu_j^2 \equiv \mu_j^2 - \mu_{j-1}^2 = 2\overline{\mu}_j\,\Delta\hat{\mu}_j, \qquad \Delta\hat{\mu}_j^2 \equiv \left|\Delta\mu_j^2\right| \tag{43}$$

we note that

$$\begin{aligned}
&\int_{\Delta\mu_j} \mu\, d\mu = \int_{\mu_{j-1}}^{\mu_j} \mu\, d\mu = \tfrac{1}{2}\Delta\mu_j^2 = \overline{\mu}_j \Delta\hat{\mu}_j && (a)\\
&\int_0^1 \hat{\mu}\, d\mu = \int_{-1}^0 \hat{\mu}\, d\mu = \sum\nolimits_{j\in D \text{ or } U} \int_{\Delta\mu_j} \hat{\mu}\, d\mu = \tfrac{1}{2}\sum\nolimits_{j\in D \text{ or } U} \Delta\hat{\mu}_j^2 = \tfrac{1}{2} && (b)
\end{aligned} \tag{44}$$

***Discretized LTE***: Both of our two discretization methods yield the discretized LTE

$$\begin{aligned}
&\overline{\mu}_f \frac{d}{dz}\tilde{I}_f(z) = -\Gamma_f(z)\tilde{I}_f(z) + \sum\nolimits_i \tilde{I}_i(z)\tilde{S}_{i\to f}(z) + \tilde{\mathcal{E}}_f(z)\\
&\hat{\overline{\mu}}_f \tilde{I}_f(z_g) = \sum\nolimits_{i\in D} \tilde{I}_i(z_g)\tilde{s}_{g,i\to f} + \tilde{\varepsilon}_{g,f} \qquad (f \in U)
\end{aligned} \tag{45}$$

where $i, f = 1,2,...,N_J$, and, using notation (5), i.e., $f(\Delta\mu) \equiv \int_{\Delta\mu} d\mu\, f(\mu)$:

$$\tilde{I}_f(z) \equiv \tilde{I}(\Delta\mu_f, z), \qquad \tilde{\mathcal{E}}_f(z) \equiv \tilde{\mathcal{E}}(\Delta\mu_f, z), \qquad \tilde{\varepsilon}_{g,f} \equiv \tilde{\varepsilon}_g(\Delta\mu_f) \tag{46}$$

As to $\Gamma_f$, $\tilde{S}_{i\to f}$ and $\tilde{s}_{g,i\to f}$, they depend on which discretization method is used.

***Method 1***: We here assume that $\tilde{I}(\mu, z)$ varies little within each sector $\Delta\mu_j$, hence is approximately equal to its *mean* value within the sector, that is:

$$\tilde{I}(\mu, z) \approx \Delta\hat{\mu}_j^{-1}\tilde{I}(\Delta\mu_j, z) \qquad \text{if} \quad \mu \in \Delta\mu_j \tag{47}$$

Thereby, $\tilde{I}(\mu, z)$ can be taken outside integrals over $\Delta\mu_j$, so that for any $f(\mu)$:

$$\int_{-1}^1 d\mu\, \tilde{I}(\mu, z) f(\mu) = \sum\nolimits_j \int_{\Delta\mu_j} d\mu\, \tilde{I}(\mu, z) f(\mu) \approx \sum\nolimits_j \tilde{I}(\Delta\mu_j, z)\Delta\hat{\mu}_j^{-1} f(\Delta\mu_j) \tag{48}$$

Integrating then the LTE (22) over $\Delta\mu_f$, and using (44) and (48), we obtain (45) with

$$\begin{aligned}
&\Gamma_j(z) = \Delta\hat{\mu}_j^{-1}\,\Gamma(\Delta\mu_j, z), \qquad \tilde{S}_{j\to f}(z) = \Delta\hat{\mu}_j^{-1}\,\tilde{S}(\Delta\mu_j \to \Delta\mu_f, z)\\
&\tilde{s}_{g,j\to f} = \Delta\hat{\mu}_j^{-1}\,\tilde{s}_g(\Delta\mu_j \to \Delta\mu_f)
\end{aligned} \tag{49}$$

where, e.g., $\tilde{s}_g(\Delta\mu_i \to \Delta\mu_f) \equiv \int_{\Delta\mu_i} d\mu_i \int_{\Delta\mu_f} d\mu_f \tilde{s}_g(\mu_i \to \mu_f)$, by (5). Obviously, (47) applies best to the diffuse radiance $\tilde{I}^d(\mu, z)$. It is *exact* if radiances are semi-isotropic, $\tilde{I}(\mu, z)_{\pm\mu>0} = \tilde{I}_\pm(z)$, as in our analytical models in Ref.[2]. Hence this method will be used in tests using these models.

***Method 2***: We here ascribe to the sharp direction $\bar{\mu}_j$ all the radiance within $\Delta\mu_j$, so that

$$\tilde{I}(\mu,z) = \sum_j \tilde{I}_j(z)\delta(\mu-\bar{\mu}_j), \qquad \tilde{I}_j(z) = \int_{\Delta\mu_j} d\mu\, \tilde{I}(\mu,z) \tag{50}$$

Substituting this into (22), and integrating over $\Delta\mu_f$, yields again (45), with this time:

$$\begin{aligned} &\Gamma_j(z) = \Gamma(\bar{\mu}_j, z), \qquad \tilde{S}_{j\to f}(z) = \tilde{S}(\bar{\mu}_j \to \Delta\mu_f, z) \\ &\tilde{s}_{g,j\to f} = \tilde{s}_g(\bar{\mu}_j \to \Delta\mu_f) \end{aligned} \tag{51}$$

In this method, photons effectively have only discrete inclinations $\bar{\mu}_j$.

In view of (6), there is little difference between (49) and (51), so that using method 1 or 2 matters little in practice. It matters only in high precision tests.[2]

***Energy balance***: In both methods 1 and 2, the vertical fluxes (30) become, either by (48) with $f(\mu) = \hat{\mu}$, or by (50):

$$\mathcal{D}(z) = \sum_{j\in D} \hat{\bar{\mu}}_j \tilde{I}_j(z), \qquad \mathcal{U}(z) = \sum_{j\in U} \hat{\bar{\mu}}_j \tilde{I}_j(z) \tag{52}$$

In particular, the incident diffuse skylight is, by (44)(b):

$$\tilde{I}^d_{j\in D}(z_0) = 2\mathcal{D}^d_0\, \Delta\hat{\mu}_j, \qquad \mathcal{D}^d_0 = \sum_{j\in D} \hat{\bar{\mu}}_j \tilde{I}^d_j(z_0) = 2\mathcal{D}^d_0 \sum_{j\in D} \hat{\bar{\mu}}_i\, \Delta\hat{\mu}_i \tag{53}$$

The absorption and emission rates (28) become

$$\mathcal{A}(z) = \sum_j \tilde{I}_j(z) A_j(z), \qquad \mathcal{E}(z) = \sum_j \tilde{\mathcal{E}}_j(z) \tag{54}$$

where, depending on the discretization method:

$$A_i(z) \equiv \left\{ \Delta\hat{\mu}_i^{-1} A(\Delta\mu_i, z) \;\text{ or }\; A(\bar{\mu}_i, z) \right\} = \Gamma_i(z) - \sum_f \tilde{S}_{i\to f}(z) \tag{55}$$

Summing the discretized LTE (45) over $f$, using (52), yields the continuity equation (29), ensuring energy balance. In particular, $\sum_f \tilde{S}_{i\to f} = \Gamma_i$ exactly if there is zero absorption.

***Sunlight***: In general, as already said, it is much better to compute accurately first-scattered sunlight, and treat that as an 'emission'. Let us now show why more explicitly. Let $\Delta\mu_{j_h}$ be the sector containing sun rays, $\mu_h \in \Delta\mu_{j_h}$. In general $\bar{\mu}_{j_h} \neq \mu_h$. Suppose now that we lump the incident sunlight together with the incident skylight within $\Delta\mu_{j_h}$, so that by (53):

$$\tilde{I}_{j_h}(z_0) = \tilde{I}(\Delta\mu_{j_h}, z_0) = 2\mathcal{D}_0^d \Delta\hat{\mu}_{j_h} + h_0 \qquad (56)$$

---

[2] Note, alternatively, that (45) also follows from assuming either $f(\mu) \approx \Delta\hat{\mu}_j^{-1} f(\Delta\mu_j)$ inside $\Delta\mu_j$, or $f(\mu) = \sum_j f(\Delta\mu_j)\delta(\mu - \bar{\mu}_j)$, for $f$ equal to $\mu$, $\Gamma$, $S$ and $s_g$. This imposes no condition on $\tilde{I}(\mu,z)$, but is an overly coarse approximation in the case $f(\mu) = \mu$.

Observe, however, that this implies, by (52), a vertical sunlight flux $\bar{\mu}_{j_h} h_0$, instead of the actual flux $\mu_h h_0$, so that the number of photons entering the canopy (per second per unit horizontal area) is altered. To maintain the proper sunlight flux, one may use, instead of (56):

$$\tilde{I}_{j_h}(z_0) = 2\mathcal{D}_0^d \Delta\hat{\bar{\mu}}_{j_h} + \left(\mu_h / \bar{\mu}_{j_h}\right) h_0 \qquad (57)$$

But still, sunlight gets attenuated as if its inclination was $\bar{\mu}_{j_h}$, due to the term $\bar{\mu}_f \, d\tilde{I}_f / dz$ on the left of (45). Also, the angles of incidence of sunrays on leaves and ground, hence $\Gamma$, $\tilde{S}$, $\tilde{s}_g$, are altered. Moreover, for consistency, the substracted sunlight in (36) should really be taken as $\tilde{I}^h(z) = (\mu_h / \bar{\mu}_{j_h}) h_0 \xi_{\bar{\mu}_{j_h}}(z)$, for the exact $\xi_{\mu_h}(z)$ might be (if $\mu_h > \bar{\mu}_{j_h}$) so much larger than $\xi_{\bar{\mu}_{j_h}}(z)$, actually *produced* by integrating the LTE (45), that $\tilde{I}^d_{j_h}(z) = \tilde{I}_{j_h}(z) - h_0 \xi_{\mu_h}(z)$ could end up negative. All this tinkering can entail sizable errors, since sunlight is so intense. This is why we prefer to treat first-scattered-sunlight, computed accurately, as an 'emission'.

The above problems do not occur in the exceptional cases that $\mu_h = \bar{\mu}_{j_h}$, if we use discretization method 2, for the coefficients (51) with $j = j_h$ are *exact* for sunlight if $\mu_h = \bar{\mu}_{j_h}$ (unlike (49) which are *averages* over $\Delta\mu_{j_h}$). We shall in fact assess how accurate our 'first-scattered-emission' treatment of sunlight is, by comparing its numerical results, for $\mu_h = \bar{\mu}_{j_h}$, with the results obtained by lumping sunlight with skylight as in (56), and using (51).

## 6 Matrix form of the LTE

We next express the discrete LTE (45) in the matrix form

$$\begin{aligned} &\frac{d}{dz} J_f(z) = -G_f(z) J_f(z) + \sum_i H_{fi}(z) J_i(z) + E_f(z) && (a) \\ &J_f(z_g) = \sum\nolimits_{i \in D} R_{g,fi} J_i(z_g) + e_{g,f} \qquad (f \in U) && (b) \end{aligned} \qquad (58)$$

We will consider two versions of (58): One where $J_j(z)$ are equal to the fluxes $\tilde{I}_j(z)$, the other where $J_j(z)$ are equal to the vertical components $\hat{\bar{\mu}}_j \tilde{I}_j(z)$ of these fluxes. We start with the latter case.

***Vertical fluxes***: Dividing (45) by $\operatorname{sgn}\bar{\mu}_f$, we obtain (58) with ($v$ stands for 'vertical'):

$$\begin{aligned} &J_j^v(z)=\hat{\bar{\mu}}_j\,\tilde{I}_j(z), \qquad G_j^v(z)=\bar{\mu}_j^{-1}\Gamma_j(z), \qquad E_f^v(z)=(\operatorname{sgn}\bar{\mu}_f)\tilde{\mathcal{E}}_f(z) \\ &H_{fi}^v(z)=(\operatorname{sgn}\bar{\mu}_f)\hat{\bar{\mu}}_i^{-1}\tilde{S}_{i\to f}(z), \qquad R_{g,fi}^v=\hat{\bar{\mu}}_i^{-1}\,\tilde{s}_{g,i\to f}, \qquad e_{g,f}^v=\tilde{\varepsilon}_{g,f} \end{aligned} \tag{59}$$

Here, $J_j^v(z)$ are vertical fluxes, and $\hat{G}_i^v,\hat{H}_{fi}^v$ are interception and scattering probabilities per unit *vertical* distance[3]. Indeed, the probability for a photon $i$ to get scattered into direction $f$ while travelling a vertical distance $|dz|$, hence an oblique distance $d\ell = dz/\bar{\mu}_i \geq 0$, is given by $\tilde{S}_{i\to f}d\ell = \tilde{S}_{i\to f}\,dz/\bar{\mu}_i = \hat{H}_{fi}^v|dz|$. The probability of absorption per unit vertical distance, and the total absorption per unit volume, are, by (54)-(55):

$$A_i^v(z)=\hat{G}_i^v(z)-\sum\nolimits_f \hat{H}_{fi}^v(z)=\hat{\bar{\mu}}_i^{-1}A_i(z), \qquad \mathcal{A}(z)=\sum\nolimits_i A_i^v(z)J_i^v(z) \tag{60}$$

(note that $G_f^v,H_{fi}^v,E_f^v$ are negative if $\bar{\mu}_f<0$ is 'up'). The total vertical photon fluxes (38) are

$$\mathcal{D}(z)=\sum\nolimits_{j\in D}J_j^v(z)=\sum\nolimits_{j\in D}J_j^{d,v}(z)+h(z)\hat{\mu}_h, \qquad \mathcal{U}(z)=\sum\nolimits_{j\in U}J_j^v(z) \tag{61}$$

The first-scattered-sunlight emission vectors are, by (41):

$$E_f^{h,v}=(\operatorname{sgn}\bar{\mu}_f)h(z)\tilde{S}(\mu_h\to\Delta\mu_f,z), \qquad e_{g,f}^{h,v}=h(z_g)\tilde{s}_g(\mu_h\to\Delta\mu_f) \tag{62}$$

***Oriented fluxes***: Let us now rather divide (45) by $\bar{\mu}_f$. This yields again (58), but with

$$\begin{aligned} &J_j(z)=\tilde{I}_j(z), \qquad G_j(z)=\bar{\mu}_j^{-1}\Gamma_j(z), \qquad E_f(z)=\bar{\mu}_f^{-1}\,\tilde{\mathcal{E}}_f(z) \\ &H_{fi}(z)=\bar{\mu}_f^{-1}\,\tilde{S}_{i\to f}(z), \qquad R_{g,fi}=\hat{\bar{\mu}}_f^{-1}\,\tilde{s}_{g,i\to f}, \qquad e_{g,f}=\hat{\bar{\mu}}_f^{-1}\tilde{\varepsilon}_{g,f} \end{aligned} \tag{63}$$

where $J_j(z)$ are now *oriented* fluxes. We then have, instead of (60)-(62):

$$A_i(z)=\sum\nolimits_f \hat{\bar{\mu}}_f\left(\hat{G}_i\delta_{fi}-\hat{H}_{fi}\right), \qquad \mathcal{A}(z)=\sum\nolimits_i A_i(z)J_i(z) \tag{64}$$

$$\mathcal{D}(z)=\sum\nolimits_{f\in D}\hat{\bar{\mu}}_f J_f(z)=\sum\nolimits_{f\in D}\hat{\bar{\mu}}_f J_f^d(z)+\hat{\mu}_h h(z), \qquad \mathcal{U}(z)=\sum\nolimits_{f\in U}\hat{\bar{\mu}}_f J_f(z) \tag{65}$$

$$E_f^h=\bar{\mu}_f^{-1}\,h(z)\tilde{S}(\mu_h\to\Delta\mu_f,z), \qquad e_{g,f}^h=\hat{\bar{\mu}}_f^{-1}h(z_g)\tilde{s}_g(\mu_h\to\Delta\mu_f) \tag{66}$$

Note that $G_j=G_j^v$ and $H_{fi}^v=\left|\bar{\mu}_f/\bar{\mu}_i\right|H_{fi}$, so that $A_i^v=\hat{G}_i-\sum_f\left|\bar{\mu}_f/\bar{\mu}_i\right|\hat{H}_{fi}$.

---

[3] Note that these probabilities per unit vertical distance may be large in grazing directions ($\bar{\mu}_i$ close to zero), with long oblique paths through $dz$.

***Discussion***: Since the matrix equation (58) represents vertical 'evolution', it is not surprising that quantities in the 'vertical' version (59) are simpler and have more direct physical meanings. Hence its use in Ref.[1]. However, if we intend to compute absorption rates by individual leaves (rather than by layers), which involve the oriented fluxes, then we may as well use the 'oriented' version (63). In any case, it does not matter much which version is used. In fact, the minuscule relative differences (less than $10^{-13}$ in our tests) between numerical results obtained using either (59) or (63) provide useful estimates of numerical accuracy. Apart from such tests, we mostly use (63) in our computations.

***Down-up block form***: Separating out the vector $\mathbf{J}(z)=\{J_j(z), j=1,2,...,N_J\}$, see (42), into subvectors $\mathbf{D}(z)=\{J_j(z), j=1,...,N_D\}$ and $\mathbf{U}(z)=\{J_j(z), j=N_D+1,...,N_J\}$ of 'down' and 'up' fluxes, we rewrite (58) in the 'down-up' block form

$$\begin{aligned}&\frac{\partial}{\partial z}\begin{pmatrix}\mathbf{D}(z)\\ \mathbf{U}(z)\end{pmatrix}=\begin{pmatrix}\mathbf{M}_{DD}(z) & \mathbf{M}_{DU}(z)\\ \mathbf{M}_{UD}(z) & \mathbf{M}_{UU}(z)\end{pmatrix}\begin{pmatrix}\mathbf{D}(z)\\ \mathbf{U}(z)\end{pmatrix}+\begin{pmatrix}\mathbf{E}_D(z)\\ \mathbf{E}_U(z)\end{pmatrix}, \qquad \mathbf{J}(z)=\begin{pmatrix}\mathbf{D}(z)\\ \mathbf{U}(z)\end{pmatrix}\\ &\mathbf{U}(z_g)=\mathbf{R}_g\mathbf{D}(z_g)+\mathbf{E}_U(z_g)\end{aligned} \tag{67}$$

where $\mathbf{M}=\mathbf{H}-\mathbf{G}$. These were our starting equations in Ref.[1].

## 7 Lambertian scattering and emission

Scattering and emission of light will henceforth be assumed purely Lambertian. Lambertian optical coefficients are, for a horizontal leaf (see section 2):

$$s(\Omega_i\to\Omega_f)=\tfrac{1}{\pi}\hat{\mu}_i\hat{\mu}_f\,\hat{\sigma}_{\operatorname{sgn}\mu_i,\,\operatorname{sgn}\mu_f}, \qquad \varepsilon(\Omega_f)=\tfrac{1}{\pi}\hat{\mu}_f\,\hat{\varepsilon}_{\operatorname{sgn}\mu_f} \tag{68}$$

where $\hat{\varepsilon}_{\pm}$ are total numbers of photons emitted (per second per unit area), and $\hat{\sigma}_{\pm\pm}$ ($\hat{\sigma}_{\pm\mp}$) are total fractions transmitted (reflected). The absorption coefficients are

$$\alpha(\Omega_i)=\hat{\alpha}_{\operatorname{sgn}\mu_i}\equiv 1-\hat{\sigma}_{\operatorname{sgn}\mu_i,+}-\hat{\sigma}_{\operatorname{sgn}\mu_i,-}, \qquad a(\Omega_i)=\hat{\mu}_i\,\hat{\alpha}_{\operatorname{sgn}\mu_i} \tag{69}$$

Similarly, Lambertian coefficients for the ground surface are

$$\begin{aligned}&s_g(\Omega_i\to\Omega_f)=\tfrac{1}{\pi}\hat{\mu}_i\,\hat{\mu}_f\,\hat{\sigma}_g, \qquad \varepsilon_g(\Omega_f)=\tfrac{1}{\pi}\,\hat{\mu}_f\,\hat{\varepsilon}_g\\ &\alpha_g(\Omega_i)=\hat{\alpha}_g=1-\hat{\sigma}_g\end{aligned} \tag{70}$$

For an inclined leaf, the coefficients may be written, using notation (1) ($u$ stands for + or -):

$$s_{\Omega_L}(\Omega_i \to \Omega_f) = \tfrac{1}{\pi}\sum\nolimits_{u_i,u_f=\pm} \hat{\sigma}_{u_i,u_f} \left|\mathbf{\Omega}_i \cdot \mathbf{\Omega}_L\right|_{u_i} \left|\mathbf{\Omega}_f \cdot \mathbf{\Omega}_L\right|_{u_f}$$
$$\varepsilon_{\Omega_L}(\Omega_f) = \tfrac{1}{\pi}\sum\nolimits_{u_f=\pm} \hat{\varepsilon}_{\Omega_L,u_f}(z) \left|\mathbf{\Omega}_f \cdot \mathbf{\Omega}_L\right|_{u_f} \tag{71}$$

where $\hat{\varepsilon}_{\Omega_L,\pm}(z)$ is allowed to depend on $\Omega_L$ and $z$, since the temperature of a leaf depends in general on its orientation and position, and so does first-scattered sunlight.

***φ–integrated coefficients***: The $\varphi_f$-integrated, $\varphi_L$-averaged coefficients (24) now become:

$$\tilde{\bar{s}}_{\mu_L}(\mu_i \to \mu_f) = 2\sum\nolimits_{u_i,u_f=\pm} \hat{\sigma}_{u_i,u_f} g_{u_i}(\mu_i,\mu_L) g_{u_f}(\mu_f,\mu_L)$$
$$\tilde{\bar{\varepsilon}}_{\mu_L}(\mu_f,z) = 2\sum\nolimits_{u_f=\pm} \bar{\bar{\varepsilon}}_{\mu_L,u_f}(z) g_{u_f}(\mu_f,\mu_L), \qquad \bar{\bar{\varepsilon}}_{\mu_L,\pm}(z) \equiv \tfrac{1}{2\pi}\int_{-\pi}^{\pi} d\varphi_L \hat{\varepsilon}_{\Omega_L,\pm}(z) \tag{72}$$

where $g_\pm$ are defined in (9). The $\varphi$–integrated ground coefficients are

$$\tilde{s}_g(\mu_i \to \mu_f) = 2\hat{\mu}_i\hat{\mu}_f\hat{\sigma}_g, \qquad \tilde{\varepsilon}_g(\mu_f) = 2\hat{\varepsilon}_g\hat{\mu}_f \tag{73}$$

***Discretized canopy coefficients***: In view of (20),(23) and (72), we have, using (5):

$$\Gamma(\Delta\mu_i,z) = \int_{-1}^{1} d\mu_L \tilde{\eta}(\mu_L,z) g(\Delta\mu_i,\mu_L)$$
$$\tilde{S}(\Delta\mu_i \to \Delta\mu_f,z) = 2\sum\nolimits_{u_i,u_f=\pm} \hat{\sigma}_{u_i,u_f} \int_{-1}^{1} d\mu_L \tilde{\eta}(\mu_L,z) g_{u_i}(\Delta\mu_i,\mu_L) g_{u_f}(\Delta\mu_f,\mu_L) \tag{74}$$
$$\tilde{\mathcal{E}}(\Delta\mu_f,z) = 2\sum\nolimits_{u_f=\pm} \bar{\bar{\varepsilon}}_{\mu_L,u_f}(z) \int_{-1}^{1} d\mu_L \tilde{\eta}(\mu_L,z) g_{u_f}(\Delta\mu_f,\mu_L)$$

The coefficients (49) (method 1) follow directly from (74). The coefficients (51) (method 2) are given by (74) but with $\Delta\mu_i$ replaced by $\bar{\mu}_i$. All these $g$ functions are known analytically (see Appendix B). The integrations over $\mu_L$ in (74) will be done numerically by putting

$$\tilde{\eta}(\mu_L,z) = \sum\nolimits_L c_L \tilde{\eta}(\bar{\mu}_L,z)\delta(\mu_L - \bar{\mu}_L) \tag{75}$$

where the discrete leaf inclinations $\bar{\mu}_L$, and the coefficients $c_L$, depend on the numerical integration rule chosen. Note that (75) corresponds to a definite physical model, with discrete leaf inclinations, for which the canopy scattering coefficients are known very accurately, whence the possibility of high precision tests.

***Discretized ground coefficients***: In view of (73), we have $\tilde{s}_{g,i\to f} = 2\hat{\bar{\mu}}_i\hat{\bar{\mu}}_f\Delta\hat{\mu}_f\hat{\sigma}_g$ in both discretization methods 1 and 2, since by (44), $\Delta\hat{\mu}_i^{-1} f(\Delta\mu_i) = f(\bar{\mu}_i) = \hat{\bar{\mu}}_i$ if $f(\mu) = \hat{\mu}$. Hence, by (63):

$$R_{g,fi} = \hat{\bar{\mu}}_f^{-1}\tilde{s}_{g,i\to f} = 2\hat{\bar{\mu}}_i\Delta\hat{\mu}_f\hat{\sigma}_g, \qquad e_{g,f} = \hat{\bar{\mu}}_f^{-1}\tilde{\varepsilon}_g(\Delta\mu_f) = 2\Delta\hat{\mu}_f\hat{\varepsilon}_g \tag{76}$$

Also, the second line of (29) becomes, using $\hat{\alpha}_g = 1 - \hat{\sigma}_g$ by (70):

$$i_g = \mathcal{D}_g - \mathcal{U}_g = \hat{\alpha}_g \mathcal{D}_g - \hat{\varepsilon}_g \qquad \text{or} \qquad \mathcal{U}_g = \hat{\sigma}_g \mathcal{D}_g + \hat{\varepsilon}_g \tag{77}$$

since $\mathcal{A}_g = \hat{\alpha}_g \mathcal{D}_g$ if $\alpha_g(\Omega_i) \equiv \hat{\alpha}_g$ (alternatively, sum (58)(b) over $f \in U$ and use (43)). If we rather work with vertical fluxes, using (59), then $R^v_{g,fi} = \Delta\hat{\mu}_f^2 \hat{\sigma}_g$ and $e^v_{g,f} = \Delta\hat{\mu}_f^2 \hat{\varepsilon}_g$.

## 8 Black leaves

'Black' leaves, which emit no light and absorb all the light they receive, provide an especially simple test case. We will use the following notations (with $z_1 < z_2$ always):

$$\begin{aligned} &\eta(z) = \int_{-1}^{1} d\mu_L\, \tilde{\eta}(\mu_L, z), \qquad \tilde{\lambda}(\mu_L, z) = \tilde{\eta}(\mu_L, z)/\eta(z) \\ &\mathcal{L}(z_1, z_2) = \int_{z_1}^{z_2} dz\, \eta(z), \qquad \mathcal{L}(z) = \mathcal{L}(z_0, z), \qquad \mathcal{L} = \mathcal{L}(z_g) = \mathcal{L}(z_0, z_g) \end{aligned} \tag{78}$$

$\eta(z)$ is the total leaf area density at $z$, while $\tilde{\lambda}(\mu_L, z)$ is the distribution of leaf inclinations. $\mathcal{L}(z_1, z_2)$ is the leaf area index (LAI) of layer $(z_1, z_2)$ (the total leaf area within the layer, per unit horizontal area). $\mathcal{L}(z)$ is the LAI above height $z$, and $\mathcal{L}$ the total LAI of the canopy.

***General solution***: With non-emitting totally absorbing black leaves, the canopy scattering and emission terms vanish, so that the LTE (22) has the solution

$$\begin{aligned} &\tilde{I}(\mu, z) = \tilde{I}(\mu, z_0)\xi_\mu(z), \qquad \xi_\mu(z) \equiv \xi_\mu(z_0, z) \\ &\xi_\mu(z_1, z_2) \equiv e^{-\Lambda_\mu(z_1, z_2)}, \qquad \Lambda_\mu(z_1, z_2) \equiv \int_{z_1}^{z_2} dz\, \Gamma(\mu, z)/\mu \end{aligned} \tag{79}$$

Observe that $\xi_\mu(z) < 1$ (attenuation) if $\mu > 0$, and $\xi_\mu(z) > 1$ (amplification) if $\mu < 0$. If the distribution $\tilde{\lambda}(\mu_L, z) \equiv \tilde{\lambda}(\mu_L)$ is uniform in $z$, then $\tilde{\eta}(\mu_L, z) = \eta(z)\tilde{\lambda}(\mu_L)$, so that by (12):

$$\begin{aligned} &\Gamma(\mu, z) = \eta(z)\gamma(\mu) \qquad \text{where} \qquad \gamma(\mu) \equiv \tfrac{1}{2\pi}\int d\Omega_L\, \tilde{\lambda}(\mu_L)\left|\mathbf{\Omega}\cdot\mathbf{\Omega}_L\right| \\ &\Lambda_\mu(z_1, z_2) = \mathcal{L}(z_1, z_2)\gamma(\mu)/\mu, \qquad \gamma(\mu) \ge 0 \end{aligned} \tag{80}$$

Now, $\gamma(\mu)$ and $\xi_\mu(z)$ can be calculated analytically for some special distributions $\tilde{\lambda}(\mu_L)$ of leaf inclinations, namely, using (8) (see also II-(32-35)):

*Horizontal* (hz): $\tilde{\lambda}(\mu_L) = \delta(\mu_L - 1), \qquad \gamma(\mu) = \hat{\mu}, \qquad \xi_\mu(z) = e^{-\mathcal{L}(z)\operatorname{sgn}\mu}$ (81)

*Semi-isotropic* (s-i): $\tilde{\lambda}(\mu_L) = \Theta(\mu), \qquad \gamma(\mu) = \tfrac{1}{2}, \qquad \xi_\mu(z) = e^{-\frac{1}{2}\mathcal{L}(z)/\mu}$ (82)

*Erect*: $\tilde{\lambda}(\mu_L) = \delta(\mu_L), \qquad \gamma(\mu) = \tfrac{2}{\pi}\sin\theta, \qquad \xi_\mu(z) = e^{-\frac{2}{\pi}\mathcal{L}(z)\sin\theta/\mu}$ (83)

(note that $\gamma$ is dimensionless, as are $\mathcal{L}$ and $\Lambda$). We see that in the semi-isotropic and erect cases, grazing 'down' light (small $\mu > 0$) hardly penetrates the canopy, while amplification of 'up' light ($\mu < 0$) blows up as $\mu \to -0$ (since oblique distances $\Delta z/\hat{\mu}$ through horizontal layers $\Delta z$ become infinite). Thus, the LAI by itself is not a reliable indicator of optical thickness.

***Transfer matrix***: As just seen above, $\xi_\mu(z)$ may blow up as $\mu \to -0$. But of course, the *discrete* photon inclinations $\hat{\bar{\mu}}_j$ have a finite minimum $\hat{\bar{\mu}}_{\min} = \hat{\bar{\mu}}_{j_{\min}}$, where $j_{\min} = N_D \pm 1$, by (42). The discrete values $\xi_{\bar{\mu}_j}(z)$, $j = 1,2,...,N_J$, are in fact elements of the transfer matrix $\mathbf{T}^{black}(z, z_0)$, which is diagonal since here there is no scattering, so that $\mathbf{H} = 0$ in (58):

$$\mathbf{T}_{ij}^{black}(z_2, z_1) = \delta_{ij}\, e^{-\Lambda_j(z_1, z_2)}, \qquad \Lambda_j(z_1, z_2) = \int_{z_1}^{z_2} dz\, \Gamma_j(z)/\bar{\mu}_j \qquad (z_2 > z_1) \tag{84}$$

If $\tilde{\lambda}(\mu_L, z) \equiv \tilde{\lambda}(\mu_L)$ is uniform in $z$, then by (80):

$$\Lambda_j(z_1, z_2) = \mathcal{L}(z_1, z_2)\gamma_j/\bar{\mu}_j, \qquad \gamma_j \equiv \gamma(\bar{\mu}_j) \geq 0 \tag{85}$$

Thus, $\hat{\Lambda}_j \leq \hat{\Lambda}_{\max} \equiv \hat{\Lambda}_{j_{\min}}$ and $\hat{T}_{ij}^{black} \leq T_{\max}^{black} \equiv e^{\hat{\Lambda}_{j_{\min}}}$ are bounded. But $T_{ij}^{black}$ can still be very large if $\mathcal{L} >> 1$, especially in grazing directions (small $\hat{\bar{\mu}}_j$). We therefore expect $\mathbf{T}$ for thick *realistic* canopies (with $\mathbf{H} \neq 0$) to likewise contain large elements (positive or negative since $\mathbf{M} = -\mathbf{G} + \mathbf{H}$ has elements of both signs), as is indeed found to be the case. This is what motivates our use of 'medium' layers in TTRG.

## Part II LAI of medium and thin layers, and numerical integration errors

This part concerns some estimates. Section 9 estimates the appropriate LAI for the 'medium' layers used in TTRG. Section 10 estimates the errors due to the numerical integration (over $z$) of first-scattered-sunlight 'propagated emissions'. Section 11 estimates appropriate LAI for the 'thin' layers in iterative integration. All these estimates are done for a specific value of $\hat{\bar{\mu}}_{\min}$, corresponding to $N_J = 18$ photon inclination sectors of $10°$ each ($N_D = N_U = 9$), so that

$$\begin{aligned} &\mu_0 = -1,\ \mu_1 = \cos 170°, ...,\ \mu_9 = 0,\ \mu_{10} = \cos 80°, ...,\ \mu_{17} = \cos 10°,\ \mu_{18} = 1 \\ &\bar{\mu}_{18} = -\bar{\mu}_1 = \tfrac{1}{2}(1 + \cos 10°) \approx 0.992, \ \ ..., \qquad \bar{\mu}_{10} = -\bar{\mu}_9 = \tfrac{1}{2}\cos 80° \approx \tfrac{1}{11.5} \end{aligned} \tag{86}$$

or $\bar{\theta}_{18} \approx 7.06°$, $\bar{\theta}_{10} \approx 85.02°$. Since $\hat{\bar{\mu}}_{\min} = \bar{\mu}_{10}$ and $\hat{\bar{\mu}}_{\max} = \bar{\mu}_{18}$, we have

$$\hat{\hat{\mu}}_{\min} \approx \tfrac{1}{11.5}\ , \qquad \hat{\hat{\mu}}_{\max} \approx 1, \qquad \left\langle \hat{\mu}^{-1} \right\rangle \equiv \tfrac{1}{\hat{\hat{\mu}}_{\max} - \hat{\hat{\mu}}_{\min}} \int_{\hat{\hat{\mu}}_{\min}}^{\hat{\hat{\mu}}_{\max}} d\mu/\mu \approx -\tfrac{1}{0.9} \ln \hat{\hat{\mu}}_{\min} \approx 2.8 \qquad (87)$$

where the mean $\left\langle \hat{\mu}^{-1} \right\rangle$ of $\hat{\mu}^{-1}$ will be needed below.

## 9 Medium layers

TTRG requires 'medium' layers whose transfer matrices are free of overly large elements. Ref.[1] gave an 'automatic' way of constructing these 'medium' layers, by successive accretion of 'thin' layers. However, it is more convenient to build them 'by hand', by estimating an appropriate LAI. This will here be done by reference to 'black' leaves.

In general, we expect transfer matrix elements, for 'medium' layers of LAI $\mathcal{L}_m$, to be bounded by $T_{\max}^{black} \sim 10^{n_m}$, where, in view of (81)-(85):

$$n_m \sim \log_{10} e^{\Lambda_{\max}} \approx \mathcal{L}_m (\log_{10} e) \gamma / \hat{\hat{\mu}}_{\min} \approx \begin{cases} \tfrac{1}{2.3} \mathcal{L}_m & (\text{horizontal (hz)}) \\ 2.5 \mathcal{L}_m & (\text{semi-isotropic (s-i)}) \end{cases} \qquad (88)$$

(since $\gamma = \hat{\mu}$ for hz, and $\hat{\hat{\mu}}_{\min} \approx \tfrac{1}{11.5}$, $\gamma = \tfrac{1}{2}$ for s-i). If the computer carries 15 digits (in double precision), then transfer matrix elements are precise to about 15 digits. Now, according to Eqs. I-(22), transmission-reflection (TR) matrices (whose elements are numbers between 0 and 1) involve *substractions* of transfer matrix elements (of size ~ $10^{n_m}$), hence are precise to roughly $15 - n_m$ digits. Thus, if we want a precision $10^{-6}$, say, then we need $15 - n_m \approx 6$, or $n_m \approx 9$, hence $\mathcal{L}_m \approx 9 \times 2.3 \approx 20$ for hz, or $\mathcal{L}_m \approx 9/2.5 \approx 3.5$ for s-i. We get nonsense if $n_m > 14$, i.e., if $\mathcal{L}_m > 32$ for hz, or $\mathcal{L}_m > 14/2.5 \approx 5.5$ for s-i. Realistic canopies tend to have s-i leaf inclinations. We indeed find that for realistic canopies of LAI beyond about 5, the *straightforward* transfer matrix method (see section I-4) yields negative values in the fluxes $\mathbf{U}_0$ reflected by the canopy. We will therefore use 'medium' layers of LAI between 1 and 3, except in numerical tests done with canopies of horizontal leaves.

## 10 Errors due to numerically integrating the propagated 'sun emissions'

If first-scattered sunlight is treated as an 'emission', then TTRG requires the 'propagated emissions' for medium layers $(z_1, z_2)$, namely

$$\mathbf{f}(z_2, z_1) = \int_{z_1}^{z_2} dz\, \mathbf{T}(z_2, z) \mathbf{E}^h(z) = \int_{\mathcal{L}_1}^{\mathcal{L}_2} d\mathcal{L}\, \mathbf{T}(\mathcal{L}_2, \mathcal{L}) \mathbf{e}_0^h\, e^{-\mathcal{L}\gamma(\mu_h)/\mu_h} \qquad (89)$$

where we assumed, for the present purposes, that $\mathbf{E}^h(z) = \eta(z) \mathbf{e}_0^h \xi_h(z)$, see (66),(23) and (85),

depends on $z$ only through $\eta(z)$ and

$$\xi_h(z) \equiv \xi_{\mu_h}(z_0, z) = e^{-\mathcal{L}(z)\gamma(\mu_h)/\mu_h} \tag{90}$$

Also, we replaced the integration variable $z$ by the dimensionless variable $\mathcal{L}(z) = \int_{z_0}^{z} dz'\eta(z')$, so that $d\mathcal{L} = \eta(z)dz$. The integral (89) is evaluated numerically using Simpson's rule, with a step size $\ell$ in $\mathcal{L}$. We here estimate the error in doing so. Because iterative integration amounts to using the trapezoid rule, we will also give in square brakets the corresponding errors. The Simpson [trapezoid] error is $\sim \frac{1}{90}\mathcal{L}_m\, \ell^4 M_4\ [\frac{1}{12}\mathcal{L}_m \ell^2 M_2]$, where $\mathcal{L}_m \equiv \left|\mathcal{L}_2 - \mathcal{L}_1\right| = \int_{z_1}^{z_2} dz\,\eta(z)$ is the LAI of the medium layer $(z_1, z_2)$, and $M_k$ is a bound on the $k$th derivative of the integrand [7,8]. Thus, the *relative* Simpson error (RSE), and the relative trapezoid error (RTE), namely the errors divided by the value of the integral, taken as roughly equal to $M_0\mathcal{L}_m$, are given by

$$RSE \sim \tfrac{1}{90}\ell^4\, M_4/M_0 \ \text{(Simpson)}, \qquad RTE \sim \tfrac{1}{12}\ell^2\, M_2/M_0 \ \text{(trapezoid)} \tag{91}$$

Now, with horizontal leaves we have $\gamma(\mu)/\mu = 1$, so that the exponentials in $\xi_h(z)$, and also in $\mathbf{T} \sim \mathbf{T}^{black}$, are of the form $e^{\pm\mathcal{L}}$, so that $M_k/M_0 \sim 1$. In the (more realistic) semi-isotropic case, $\gamma(\mu)/\mu = \frac{1}{2}\mu^{-1}$, so that $\xi_h(z)$ if $\mu_h \geq \frac{1}{2}$, and also $\mathbf{T}$ for the dominant non-grazing directions, again behave like $e^{\pm\mathcal{L}}$, hence $M_k/M_0 \sim 1$. But for small $\mu_h$, the $z$ dependence in the s-i case is dominated by the rapidly decreasing $\xi_h(z)$, so that $M_k/M_0 \sim (\frac{1}{2}\mu_h^{-1})^k$, by (90) with $\gamma = \frac{1}{2}$. So in summary we may put:

$$\begin{aligned} &RSE \sim \tfrac{1}{90}\ell^4 \quad \left[RTE \sim \tfrac{1}{12}\ell^2\right] \quad \text{(horizontal leaves, and s-i if } \mu_h \geq \tfrac{1}{2}) \\ &RSE \sim \tfrac{1}{90}(\tfrac{1}{2}\ell/\mu_h)^4 \quad \left[RTE \sim \tfrac{1}{12}(\tfrac{1}{2}\ell/\mu_h)^2\right] \quad \text{(s-i leaves if } \mu_h < \tfrac{1}{2}) \end{aligned} \tag{92}$$

## 11 Thin layers in iterative integration

Iterative integration (see section I-12) is done using $N$ 'thin' layers of LAI $\ell$. The transmission-reflection matrices and emission vectors for these thin layers are treated to first order in $\ell$. The resulting errors, and acceptable values of $\ell$, will now be estimated.

Consider a vertically uniform canopy of total LAI $\mathcal{L}$, so that $\ell \equiv \mathcal{L}/N$. Treating the iteration equations I-(47) to first order in $\ell$ amounts to assuming that a photon scattered or emitted by a leaf element inside a thin layer has negligable probability of hitting another leaf before exiting the thin layer. This probability is $\sim(\frac{1}{2}\ell)\langle\gamma/\hat{\mu}\rangle$ (since the mean *vertical* distance out is $\frac{1}{2}\ell$). Putting this equal to $0.1$ say yields $\ell \sim 0.2$ for horizontal leaves ($\gamma = \hat{\mu}$), and $\frac{1}{4}\ell\langle\hat{\mu}^{-1}\rangle \approx \frac{2.8}{4}\ell \approx 0.1$, hence $\ell \sim 0.1$, for semi-isotropic leaves $\left(\gamma = \frac{1}{2}\right.$, and $\langle\hat{\mu}^{-1}\rangle \approx 2.8$ by (87)$\left.\right)$. Now, as was argued in section I-12, when treated to first order in $\ell$, transmission matrices are too small, and reflection matrices too large. Thus, since in realistic canopies photons undergo numerous back and forth transmissions and reflections through the various thin layers, single thin layer errors are of *both* signs, hence do not accumulate. Indeed, computer experiments on 'realistic' canopies (with s-i leaves) indicate that using $\ell \approx 0.1$ is usually adequate (the precision of iterative integration is here inferred by comparing with TTRG results, or by examining convergence as $\ell \to 0$). For instance, for an s-i canopy with $\mathcal{L} = 10$, the relative errors in the radiances at the bottom $z_g$ of the canopy are between 1 and 5% if $\ell = 0.1$, and about 10 times less if $\ell = 0.01$ (the error seems roughly linear in $\ell$). However, errors are much larger in the case of grazing sunlight (very small $\mu_h$) [4], and are roughly as given inside the square brackets in (92), since iterative integration in effect numerically integrates the first-scattered sunlight 'emissions' with the trapezoid rule. With $\mu_h = 0.01$ for instance, results are way off if $\ell = 0.1$, and err by 0.5 to 2% if $\ell = 0.01$.

In some artificial situations, however, thin layer errors do accumulate, for instance in the case of black leaves, where there are no back and forth reflections:

***Black leaves***: Denote $\lambda_j \equiv \Lambda_j / N$, where $\Lambda_j \equiv \Lambda_j(z_0, z_g)$ is for the whole canopy. With black leaves, the canopy reflection matrices $\mathbf{r}, \boldsymbol{\rho}$ are zero, while the transmission matrices $\mathbf{t}, \boldsymbol{\tau}$ are just the transfer matrices. Applying (84) to thin layers $(z_1, z_2)$, we get iteration equations

$$\mathbf{D}_n = \mathbf{t}\,\mathbf{D}_{n-1}, \quad \mathbf{U}_N = \mathbf{R}_g\mathbf{D}_N, \quad \mathbf{U}_{n-1} = \boldsymbol{\tau}\mathbf{U}_n, \quad t_{ij} = \tau_{ij} = \delta_{ij}\, e^{-\lambda_j} \tag{93}$$

(see section I-12) where

$$\lambda_j \equiv \Lambda_j / N = \ell\gamma_j / \bar{\mu}_j, \quad \Lambda_j \equiv \Lambda_j(z_0, z_g), \quad \ell \equiv \mathcal{L}/N \tag{94}$$

Thus, $\mathbf{D}_N = \mathbf{t}^N\mathbf{D}_0$ and $\mathbf{U}_0 = \boldsymbol{\tau}^N\mathbf{U}_N$, that is, $D_j(z_g) = e^{-\Lambda_j}D_j(z_0)$ and $U_j(z_0) = e^{-\Lambda_j}U_j(z_g)$, as in (79). Here, $\mathbf{t}$ and $\boldsymbol{\tau}$ are exact. We now estimate the error done in treating $\mathbf{t}$ and $\boldsymbol{\tau}$ in (93) to first order in $\ell$. For a single thin layer, approximating $e^{\lambda} = 1 + \lambda + \frac{1}{2}\lambda^2 + ... \approx 1 + \lambda$ implies an error $\delta_1 \approx \frac{1}{2}\lambda^2$ (we omit subscripts *j*). However, single layer errors *accumulate*. Indeed, treating (93) to first order in $\ell$ amounts to approximating $e^{-\Lambda} = e^{-N\lambda} \approx (1-\lambda)^N$, resulting in a *relative* error (i.e., the error divided by the radiance $e^{-\Lambda}\mathcal{D}_0$) given by, in view of (A.1):

$$\delta_2 \equiv \left| e^{-\Lambda} - (1-\lambda)^N \right| e^{\Lambda} \approx \tfrac{1}{2}\Lambda^2/N = \tfrac{1}{2}\lambda\Lambda = \tfrac{1}{2}N\lambda^2 = N\delta_1 \qquad (\lambda = \Lambda/N) \tag{95}$$

---

[4] Note that in general the weight of small $\hat{\mu}$ (excluding $\mu_h$) is small on average, because most of the diffuse light reaching down travelled nearly vertically (along paths of shortest $\Delta z/\hat{\mu}$), after having been scattered by leaves higher up.

Let $N_z$ be the number of thin layers above height $z$. Now it may happen that when $N_z$ is large enough that $\delta_2(z) = \frac{1}{2}N_z\lambda^2$ is sizable (for a given $\lambda$), the attenuation $e^{-\Lambda} \equiv e^{-N_z\lambda}$ is already so strong (hence radiances so small) that errors are irrelevant. So what really matters here is the error relative to the *incident* flux $\mathcal{D}_0$ (not relative to $e^{-\Lambda}\mathcal{D}_0$ as in (95)), namely $\delta_3 \equiv \delta_2 e^{-\Lambda} = \frac{1}{2}\lambda\Lambda e^{-\Lambda}$. Since $(d/d\Lambda)\Lambda e^{-\Lambda} = (1-\Lambda)e^{-\Lambda}$, the maximum of $\Lambda e^{-\Lambda}$ occurs at $\Lambda = 1$, and is $1/e$. Hence

$$\delta_3 \approx \lambda/2e = (\ell/2e)(\gamma/\hat{\mu}) \approx \tfrac{1}{5.4}\ell\gamma/\hat{\mu} \tag{96}$$

For semi-isotropic leaves, $\gamma = \frac{1}{2}$, so that $\delta_3 \approx \frac{1}{4}e^{-1}\,\ell/\hat{\mu} \approx \frac{1}{11}\ell/\hat{\mu}$ ranges from $\delta_3 \approx \frac{1}{11}\ell$ (for $\hat{\mu} = 1$) to $\delta_3 \approx \ell$ (for $\hat{\mu} = \hat{\mu}_{\min} = \frac{1}{11.5}$), with a mean $\langle\delta_3\rangle \approx \frac{1}{11}\ell\langle\hat{\mu}^{-1}\rangle \approx \frac{1}{4}\ell$, by (87). For horizontal leaves, $\gamma = \hat{\mu}$, so that $\delta_3 \approx \frac{1}{5.4}\ell$. So one may take, in general, $\delta_3 \sim \frac{1}{4}\ell$ (for black leaves). For instance, $\ell \approx 0.1$ implies $\delta_3 \sim \frac{1}{4}\ell \approx 2.5\%$.[5]

*Remark*: We will encounter in section 14 light trapping situations where radiances *increase* down the canopy, and errors again *accumulate*, so that the relevant error is given by $\delta_2$, and much smaller $\ell$ must be used. For instance, (94),(95) with $\gamma = \frac{1}{2}$ yield $\delta_2 \approx \frac{1}{8}\mathcal{L}\ell/\hat{\mu}^2 > \frac{1}{8}\mathcal{L}\ell$ (since $\hat{\mu} \le 1$), so that if $\mathcal{L} = 8$, say, and we use $\ell \approx 0.1$, then $\delta_2 > 10\%$.

---

[5] We should check that using $\delta_2 \approx \frac{1}{2}\Lambda^2 N^{-1}$ is valid for $\Lambda = 1$ and the values of $N = \Lambda/\lambda$ involved here. With $\gamma = \frac{1}{2}$, $\ell \approx 0.1$, we have $N = \lambda^{-1} = \gamma/\ell\hat{\mu} = 5\hat{\mu}^{-1}$, so that $5 < N < 57$ since $\frac{1}{11.5} < \hat{\mu} < 1$. For $N > 5$, the correction $N^{-2}\left(\frac{1}{8} - \frac{1}{3}\right) = -\frac{5}{24}N^{-2}$ in (A.1) (with $x = \Lambda = 1$) is indeed sufficiently smaller than $\delta_2 \approx \frac{1}{2}\Lambda^2/N = \frac{1}{2}N^{-1}$, their ratio being $\frac{10}{24}N^{-1} < \frac{1}{12}$.

## Part III Numerical tests

Section 12 describes our computational setup, and gives the results of some general numerical tests. Section 13 presents tests done with analytic models producing isotropic radiances. Section 14 presents tests done with horizontal leaves, in situations of extreme light trapping. Finally, section 15 concerns iterative integration.

## 12 Computational setup and general tests

All our computations carry 15 digits (double precision).

Leaves and ground are always Lambertian.

As in (86), we use $N_J = 18$ photon inclination sectors $\Delta\mu_j$, of equal angular sizes $\Delta\theta_j = 10°$. We integrate numerically over leaf inclinations $\mu_L$, using (75) adapted to the midpoint rule, with the range $(0,1)$ of leaf inclinations partitioned into equal intervals $\Delta\mu_L$:

$$\Delta\mu_L = (\mu_{L1}, \mu_{L2}), \qquad \bar{\mu}_L = \tfrac{1}{2}(\mu_{L1} + \mu_{L2}), \qquad c_L = \Delta\hat{\mu}_L \equiv |\mu_{L1} - \mu_{L2}| \tag{97}$$

We use $9$ equal intervals $\Delta\mu_L$ of sizes $\Delta\hat{\mu}_L = \frac{1}{9}$. Horizontal (erect) leaves require a single value of $L$ in (75), with $\bar{\mu}_L = 1$ ($\bar{\mu}_L = 0$).[6]

***Integration of propagated emissions***: We subdivide each medium layer $(z_m, z_{m+1})$ into an *even* number $N_m$ of thin layers, in order to integrate the propagated emissions $\mathbf{f}^{(m)} \equiv \mathbf{f}(z_{m+1}, z_m)$ numerically using Simpson's rule. However, to determine the radiances *between* thin layers (inside the medium layer), we require $\mathbf{f}(z_{m+n}, z_m)$ for $n = 1,2,...,N_m$, where $z_{m+N_m} = z_{m+1}$. For $n$ odd, the integration over the last (odd) thin layer $(z_{m+n-1}, z_{m+n})$ is not as accurate as the integration over *pairs* of layers, as explained in Appendix C, so that the precisions of the computed radiances between thin layers alternate by about one digit (more precise for $n$ even, less for $n$ odd).

***Discretization methods* 1 *and* 2**: In section 5, we gave two different methods (1 and 2) for discretizing the photon inclinations $\mu$. In general, radiances computed using these two methods differ by between $10^{-4}$ and $10^{-3}$.

---

[6] Note that for (semi)isotropic leaf inclinations, we do not use the analytic canopy optical coefficients obtained in section 9 of Ref. [2], because we are unable to integrate those analytically over $\Delta\mu_f$. We prefer to use (75) to ensure precise energy balance.

***Working with radiances or with vertical fluxes***: One can choose to work with radiances $\tilde{I}(\Delta\mu_j, z)$ by using the LTE (58) with (63), or with vertical fluxes $\hat{\bar{\mu}}_j \tilde{I}(\Delta\mu_j, z)$ by using (58) with (59). Radiances computed in these two ways generally differ by less than $10^{-13}$.

***Sunlight***: As discussed in section 5, sunlight may be treated in two different ways:

(i) '*Incident' method*: Lump sunlight with the skylight incident in sector $\Delta\mu_{j_h}$, using (57).

(ii) '*Emission' method*: Treat first-scattered sunlight as an 'emission'.

To compare these two methods, for 'realistic' canopies (semi-isotropic leaves), we assume that *only* sunlight is incident, and use (51) (discretization method 2). We first let $\mu_h = \bar{\mu}_{j_h}$ to make the 'incident' method *exact*, so that the relative differences (r.d.) between results yielded by the 'incident' and 'emission' methods are due solely to Simpson numerical integration errors in the 'emission' method, that is, r.d ~ RSE in Eq. (92). The actual errors roughly agree with (92) (within factors 10 or so). For instance, with a total canopy LAI $\mathcal{L} = 10$ and $\mu_h > 0.3$, we get $10^{-6} < \text{r.d.} < 10^{-5}$ if $\ell = 0.1$, and $10^{-10} < \text{r.d.} < 10^{-9}$ if $\ell = 0.01$ (reflecting the $\ell^4$ dependence). With $\mu_h \approx 0.05$ and $\ell = 0.1$ we get $\text{r.d.} \sim 10^{-3}$.

We next do calculations with $\mu_h \neq \bar{\mu}_{j_h}$. Radiances yielded by the 'incident' and 'emission' methods then differ by up to 20 percent, giving the error due to nudging the sunlight inclination from $\mu_h$ to $\bar{\mu}_{j_h}$ in the 'incident' method. It is not easy to estimate theoretically this 'nudging' error, because much of the light that penetrates deeper (especially with grazing sunlight) is sunlight scattered nearly *vertically* downwards (minimum $\Delta z/\hat{\mu}$) by leaves higher up. But the general tendency is as expected: Too much (not enough) light reaches down if $\mu_h$ is nudged towards vertical (horizontal), since this decreases (increases) $\mathcal{L}/\mu_h$. If in the 'incident' method we use (56) rather than (57), thereby altering the *flow* of energy into the canopy, then the discrepancy with 'emission' results more than doubles, and may reach above $200\%$ at very small $\mu_h$ (e.g., more than $400\%$ for $\mu_h = 0.01$, in the case of visible light). The 'emission' method is clearly the more accurate and reliable (unless $\mu_h \approx \bar{\mu}_{j_h}$ by chance), and is the one we use in practice.

## 13 Non-absorbing leaves and ground

Let us assume that: (i) leaf area densities are $\varphi$-symmetric; (ii) incident light is semi-isotropic; and (iii) leaves and ground are non-absorbing and non-emitting. Then, in view of Eq.II-(89) with $c_{\pm}=0$, the radiances inside the canopy are isotropic, $\tilde{I}(\mu,z)=\tilde{I}(z)$, with

$$\tilde{I}(z)=\tilde{I}(z_0)e^{\frac{1}{2}\mathcal{L}(z)\langle\mu_L\rangle(b_+-b_-)}=\tilde{I}(z_0)e^{\mathcal{L}(z)\langle\mu_L\rangle\Delta\hat{\sigma}}, \qquad \Delta\hat{\sigma}\equiv\hat{\sigma}_{++}-\hat{\sigma}_{--} \tag{98}$$

where $b_+=\hat{\sigma}_{++}+\hat{\sigma}_{-+}$, $b_-=\hat{\sigma}_{--}+\hat{\sigma}_{+-}$, and $\langle\mu_L\rangle$ is the mean leaf inclination. We noted that $b_+-b_-=2\Delta\hat{\sigma}$ since zero absorption implies $\hat{\sigma}_{-+}=1-\hat{\sigma}_{--}$ and $\hat{\sigma}_{+-}=1-\hat{\sigma}_{++}$. Observe that $\tilde{I}(z)$ increases down the canopy if $\Delta\hat{\sigma}>0$, implying light trapping.

We here discretize photon inclinations using (49) (discretization method 1), since it is *exact* for isotropic radiances. Note that using discrete leaf inclinations entails no loss of precision (since these are still $\varphi$-symmetric). Leaf inclinations are taken semi-isotropic, so that $\langle\mu_L\rangle=\frac{1}{2}$.

As argued in section 9, TTRG computations loose $n_m$ digits of precision due to the thickness $\mathcal{L}_m$ of medium layers. Other sources of error are not easy to estimate, but in the light trapping case $\Delta\hat{\sigma}>0$, analogy to the simpler case of horizontal leaves, treated in section 14 below, suggests an additional loss of $n_I$ digits in precision, where $n_I\approx\log I(z_g)=(\log e)\frac{1}{2}\mathcal{L}\Delta\hat{\sigma}$ $\approx 0.2\mathcal{L}\Delta\hat{\sigma}$, if $\langle\mu_L\rangle=\frac{1}{2}$. Thus, the expected precision is $n_a$ digits, where, using (88):

$$n_a\approx 15-n_m-n_I\approx 15-2.5\mathcal{L}_m-0.2\mathcal{L}\Delta\hat{\sigma} \tag{99}$$

This is roughly what we get for the cases we treated, with $\mathcal{L}$ ranging from 1 to 30, and $\mathcal{L}_m$ between 1 and 3. For instance, if $\Delta\hat{\sigma}=1$, $\mathcal{L}=30$, and $\mathcal{L}_m=3$, then (99) yields $n_a\approx 1$, while TTRG numerical results differ by $\sim 5\%$ from the analytic radiances.

If we obtain the radiances between *medium* layer by iterating Eqs. I-30 (rather than using the Green's matrix as in TTRG), then the results agree to $15-n_I$ digits with TTRG, since both computations use the same transmission-reflection matrices, with the same error $n_m$.

Computed radiances turn out to be isotropic to better than $10^{-10}$. If we add in emissions given by Eqs. II-(86), then precision is better than $10^{-3}$ in all the cases we treated. If we rather use discretization method 2 (i.e., (51) instead of (49)), then precision is $\sim 10^{-3}$ ( as expected since methods 1 and 2 yield results differing by $\sim 10^{-3}$ in general).

## 14 Horizontal leaves, extreme light trapping

In this section, leaves are horizontal. Methods 1 and 2 for discretizing photon inclinations are then equivalent, and yield canopy optical coefficients similar to the ground coefficients (76). Because the 'effective' LAI does not increase in oblique photon directions, we expect higher precisions than with semi-isotropic leaves (for a given LAI). Also, we can use thicker 'medium' layers, as was argued in section 9. We use $\mathcal{L}_m \le 10$, so that $n_m \le 4$ by (88).

With semi-isotropic incident radiances, radiances throughout the canopy should be semi-isotropic. TTRG computed radiances are semi-isotropic to better than $10^{-10}$ (for $\mathcal{L} = 10$).

***Black leaves***: With non-absorbing non-emitting black leaves, 'down' radiance at $z = z_g$ should be $e^{-\mathcal{L}} D_0$, and reflected light at $z = z_0$ should be zero if the ground is totally absorbing ($R_g = 0$), and $e^{-2\mathcal{L}} D_0$ if totally reflecting ($R_g = 1$). TTRG computations yield these analytic values to better than $10^{-11}$, for $\mathcal{L} = 1$ to 30, even if we use a single 'medium' layer (of LAI up to 30). Indeed, since here $\mathbf{T}_{UD} = \mathbf{T}_{DU} = 0$, the transmission-reflection matrices I-(22) involve no substractions of (large) transfer matrix elements.

***Extreme light trapping***: This is produced by a totally reflective ground, and leaves which totally transmit 'down' light, but totally reflect 'up' light. We showed in section II-12 that with only semi-isotropic skylight incident, the fluxes at the bottom $z = z_g$ of the canopy are

$$D_g = U_g = \left(1 + r + r^2 + ...\right) D_0 = \left(1 - r\right)^{-1} D_0 = \tau^{-1} D_0, \qquad r \equiv 1 - \tau, \qquad \tau \equiv e^{-\mathcal{L}} \tag{100}$$

Here, $D$ stands for the vector $(D, D, ..., D)$, i.e., the $D_j = D$ are all equal. Also, $r$ stands for $r$ times the $9 \times 9$ unit matrix (if $N_D = N_U = 9$).

***TTRG calculation***: In view of Eq.II-(108), we here have, if we use a *single* medium layer:

$$\mathcal{Q} = \begin{pmatrix} 0 & r \\ 1 & 0 \end{pmatrix} \quad , \quad \begin{pmatrix} D_g \\ U_g \end{pmatrix} = \mathcal{G}\mathcal{E} = \frac{1}{1-r} \begin{pmatrix} 1 & r \\ 1 & 1 \end{pmatrix} \begin{pmatrix} D_0 \\ 0 \end{pmatrix} = \frac{1}{1-r} \begin{pmatrix} D_0 \\ D_0 \end{pmatrix} \tag{101}$$

where we used $\mathcal{E} = (D_0, 0)$ (no 'true' emissions), and noted that $\mathcal{Q}^2 = r$, so that

$$\mathcal{G} = (1 - \mathcal{Q})^{-1} = 1 + \mathcal{Q} + \mathcal{Q}^2 + ... = (1 - r)^{-1}(1 + \mathcal{Q}) \tag{102}$$

Thus, TTRG yields (100) with $\tau^{-1}$ computed as $\tau^{-1} = (1 - r)^{-1}$. Now, $\tau$ and $r$ are accurate to about 14 digits. So if $\tau$ is very small, then $r = 1 - \tau \approx 1$, so that the small number $\tau = 1 - r$ is the difference of two 'large' numbers (1 and $r \approx 1$), hence is precise to $14 - n_\tau$ digits, where $n_\tau \sim -\log_{10}\tau = \mathcal{L}\log_{10} e \approx 0.43\mathcal{L}$. Hence we expect computed radiances precise to $14 - n_\tau$ digits. This is roughly the case: For instance, if $\mathcal{L} = 10$, then $e^{\mathcal{L}} = 22026$, and we get 10 digits precision; if $\mathcal{L} = 30$, then $e^{\mathcal{L}} \approx 1.0686 \times 10^{13}$, and precision is down to about $3\%$. We get nonsense once $\mathcal{L} > 14\ln 10 \approx 32$ (recall that $\ln 10 = 1/\log_{10} e$). But of course, as said after Eq.II-(109), actual TTRG computations would then use *several* medium layers (of LAI $\mathcal{L}_m < 10$ say), in order that no $r_m = 1 - \tau_m$ be too close to 1.

***Sunlight***: If *only* sunlight is incident, and is treated as an 'emission', then as shown in section II-12, we have again $D_g = (1-r)^{-1} D_0$. The computed $D_g$ agrees to $\sim 10^{-6}$ with that computed above (with *only* skylight incident), for *all* values of $\mathcal{L}$ treated, between 1 and 30, and with $\ell = 0.1$, in line with the relative Simpson error $\sim \frac{1}{90}\ell^4 \approx 10^{-6}$ in (92).

***Iteration***: Rather than use the Green's matrix, we can obtain the radiances between medium layers by iterating Eqs. I-(30). Consider the case of a single medium layer (the whole canopy). The iteration equations are then $D_g^{(k)} = D_0 + rU_g^{(k-1)}$, $U_g^{(k)} = D_g^{(k)}$, $U_0^{(k)} = \tau U_g^{(k)}$, so that

$$D_g^{(k+1)} = D_0 + rD_g^{(k)}, \qquad r = 1 - \tau \tag{103}$$

where $D$ again stands for the vector $(D, D, ..., D)$. Iteration stops when $D_g^{(k)}$ is large enough that the loss $D_g^{(k)} - rD_g^{(k)} = \tau D_g^{(k)}$ is (nearly) equal to the gain $D_0$, i.e., when $D_g^{(k)} \approx \tau^{-1} D_0$. In view of Eq.II-(103), the number of iterations needed to obtain a relative accuracy $\delta_r$ is $K \approx -\tau^{-1}\ln\delta_r$. Now, $r$ is either too large or too small by $\varepsilon \sim 10^{-15}$. So multiplication by $r$, repeated $K$ times (so that roundoff errors accumulate), causes a relative error $\left|(r \pm \varepsilon)^K - r^K\right| / r^K \approx K\varepsilon/r \approx K\varepsilon \sim |\ln\delta_r|\tau^{-1} \times 10^{-15} \approx |\ln\delta_r| \times 10^{-15+n_\tau}$ (since $r \approx 1$), so that precision is $\sim 14 - n_\tau$ digits if $|\ln\delta_r| \le 10$ (i.e., if $\delta_r \ge 10^{-10}$).

## 15 Iterative integration

We now discuss iterative integration applied to the above extreme light trapping situation. Let us repeat the discussion in the last paragraph of section II-12, but with no sunlight present, and putting $\Delta\tau = 1 - \ell$ explicitly. We have, for the total vertical fluxes:

$$\mathcal{D}_n = \mathcal{D}_{n-1} + \ell\mathcal{D}_n, \qquad \mathcal{U}_N = \mathcal{D}_N, \qquad \mathcal{U}_{n-1} = (1-\ell)\mathcal{U}_n \qquad \ell \equiv \mathcal{L}/N \tag{104}$$

Thus, $\mathcal{U}_n = (1-\ell)^{N-n}\mathcal{U}_N$ (where $\mathcal{U}_N = \mathcal{U}_g$), and the $k$th iterate is, using II-(100):

$$\mathcal{U}_N^{(k)} = \mathcal{D}_N^{(k)} = \mathcal{D}_0 + \ell \sum\nolimits_{n=0}^{N} \mathcal{U}_n^{(k-1)} = \mathcal{D}_0 + r_\ell \mathcal{U}_N^{(k-1)}, \qquad r_\ell \equiv 1-\tau_\ell, \quad \tau_\ell \equiv (1-\ell)^N \tag{105}$$

Hence, putting $\mathcal{U}_N^{(0)} = 0$ , we get $\mathcal{U}_N^{(1)} = \mathcal{D}_0 + r_\ell \mathcal{D}_0$, $\mathcal{U}_N^{(2)} = (1 + r_\ell + r_\ell^2)\mathcal{D}_0$,...,

$$\mathcal{U}_N^{(k)} = \sum\nolimits_{m=0}^{k} r_\ell^{\,m} \mathcal{D}_0 = (1-r_\ell)^{-1}(1-r_\ell^{k+1})\mathcal{D}_0 \tag{106}$$

So as $k \to \infty$, $r_\ell^{k+1} \to 0$, and $\mathcal{U}_g^{(k)} / \mathcal{D}_0$ converges from *below* to $(1-r_\ell)^{-1} = \tau_\ell^{-1} = \left(1-\mathcal{L}/N\right)^{-N}$. And the latter converges from *above* to $e^{\mathcal{L}}$ as $N \to \infty$ (indeed, compare, e.g., for $\mathcal{L} = 10$, the value $e^{\mathcal{L}} \approx 22\,026$, with $\left(1-\mathcal{L}/N\right)^{-N} \approx 37\,648$, and also $\left(1+\mathcal{L}/N\right)^{N} \approx 13\,780$, if $N = 100$).

Let $\mathbf{J}_n^{(k)} = \left(\mathbf{D}_n^{(k)}, \mathbf{U}_n^{(k)}\right)$ be the fluxes obtained in the *k*th iteration of equations (104). Our computer code asks that iteration stops when

$$\left(J_{n,j}^{(k)} - J_{n,j}^{(k-1)}\right) \Big/ J_{n,j}^{(k-1)} < \delta_i \qquad \text{(for all } n, j\text{)} \tag{107}$$

i.e., when stability to within $\delta_i$ is attained. If the incident radiance is semi-isotropic, then so are the radiances inside the canopy, so that (107) is equivalent to $\left(\mathcal{U}_n^{(k)} - \mathcal{U}_n^{(k-1)}\right) \big/ \mathcal{U}_n^{(k-1)} < \delta_i$. In the present case, $n = N$ (at which $\mathcal{U}$ is largest) is determinant, so that in view of (100), iteration stops when $\left(s_k - s_{k-1}\right) \approx \delta_i s_{k-1} = \delta_i (1-r^k)/(1-r)$, where $s_k \equiv 1 + r + r^2 + ... + r^k$. Since $\left(s_k - s_{k-1}\right) = r^k$, we can solve for $r^k$, to get $r^k(\delta_i + \tau) = \delta_i$ (since $r = 1-\tau$), whence:

$$k = -\ln\left(1 + \tau/\delta_i\right) \big/ \ln(1-\tau) \approx \tau^{-1} \ln\left(1 + \tau/\delta_i\right) \qquad (\text{for } \tau << 1) \tag{108}$$

The actual number of iterations is very close to the theoretical value (108) (e.g., with $\mathcal{L} = 10$ and $\delta_i = 10^{-11}$, the number $k$ predicted by (108) is 322 357, while the actual one is 322 358).

Now, in view of Eq.II-(103), the number $k$ of iterations needed to approximate the analytic value $\mathcal{U}_N = \tau^{-1}\mathcal{D}_0$ with a relative error $\delta_r$ is given by $k \approx -\tau^{-1}(\ln \delta_r)$. It follows from (108) that $\ln \delta_r \approx -\ln\left(1 + \tau/\delta_i\right)$, hence $1 + \tau/\delta_i \approx \delta_r^{-1}$, or $\delta_i \approx \tau \delta_r$ if $\delta_r << 1$. Thus, if $\tau << 1$, then $\delta_i$ must be taken *much smaller* than the desired relative error $\delta_r$ (showing that $\delta_i$ can sometimes be quite unrelated to $\delta_r$). The trouble here is that because $r^{k+m}$ decreases very slowly as $m$ increases (since $r$ is very close to 1), many terms of size comparable to $r^k$ add up. A better convergence criterion would here be $r^k = s_k - s_{k-1} < \delta_r$, or $\left(\mathcal{U}_g^{(k)} - \mathcal{U}_g^{(k-1)}\right) \big/ \mathcal{D}_0 < \delta_r$. But this would not do if above our light trapping layer (LTY) we had another thick layer of totally absorbing horizontal leaves, decreasing the flux $\mathcal{D}_0$ incident on the LTY by a factor $\tau_0 << 1$, so that a proper criterion would then rather be $\mathcal{D}^{(k)} - \mathcal{D}^{(k-1)} < \mathcal{D}_0\, \tau_0 \delta_r$. So just specifying a convergence criterion can, in certain cases, be problematic.

***Error due to treating*** $\mathbf{t}, \mathbf{r}, \boldsymbol{\rho}, \boldsymbol{\tau}$ ***to first order in*** $\ell$: As shown in section I-12, when treated to first order in $\ell$, transmissions are too small and reflections too large, which means too much trapping in the present case. For instance, if $\mathcal{L} = 10$ and we ask that $\delta_2 = 0.01 = 1\%$, see (95), then we need $N \approx 5000$ hence $\ell = \mathcal{L}/N \approx 0.002$. If we were to use the 'usual' $\ell = 0.1$, hence $N = 100$, then we would get $\left[(1 - \mathcal{L}/N)^{-N} - e^{\mathcal{L}}\right]/e^{\mathcal{L}} \approx 0.71 = 71\%$ ! [7]

We thus see that in the iterative integration method, one should really check the stability of results against both $\delta_i$ and $N$, that is, vary these numbers until stable (presumably correct) values are attained.

***Computation times***: For a relative accuracy $\delta_r$, we need $k \approx -\tau^{-1} \ln \delta_r$ iterations. The number of thin layers needed in order that $\left(1 - \mathcal{L}/N\right)^{-N}$ be equal to $e^{\mathcal{L}}$ to a relative accuracy $\delta_2 \approx \delta_r$ is $N \approx \frac{1}{2}\mathcal{L}^2/\delta_r$, by (95) with $\Lambda = \mathcal{L}$. The computation time is thus

$$CP \approx sNk \approx -s\left(\tau^{-1} \ln \delta_r\right)\left(\tfrac{1}{2}\mathcal{L}^2/\delta_r\right) \approx s\,\mathcal{L}^2 e^{\mathcal{L}} \delta_r^{-1} \log_{10} \delta_r^{-1}$$

where we put $\frac{1}{2}\ln 10 \approx 1.15 \approx 1$, and $s$ is the number of seconds required per iteration and per layer. Our various computations (using a 1 GHz Pentium IV) indicate that $s \sim 2 \times 10^{-5}$. For instance, with $\mathcal{L} = 10$ and $\delta_r = 10^{-2}$, we get $k \approx -e^{10} \ln 10^{-2} \approx 10^5$, $N \approx \frac{1}{2}\mathcal{L}^2/\delta_r = 5000$, so that $CP \approx sNk \approx 10\,000$ seconds $\approx 2.8$ hours (roughly what we get).

In all our tests, TTRG computations times are less than a second (including the computation of all matrices).[8]

---

[7] Note that in this case (95) (of first order in $N^{-1}$) yields $\delta_2 \approx \frac{1}{2}\mathcal{L}^2/N = \frac{1}{2}$, while (A.1) to second order in $N^{-1}$ yields $\delta_2 \approx \frac{1}{2} + \left(\frac{1}{3}10^3 + \frac{1}{8}10^4\right)10^{-4} \approx 0.66$.

[8] To compute $\mathbf{A}^{-1}\mathbf{B}$, we use what MATLAB calls 'left division' **A\B**, based on Gauss elimination algorithms. This is more accurate than calculating $\mathbf{A}^{-1}$ separately, and then multiplying into **B**.

## 16 Conclusion

All the numerical tests discussed in the preceding sections indicate that TTRG numerical integration of the light transport equation (LTE) is highly accurate and rapid, even in cases of extreme light trapping where iterative integration is hardly practical.

One might object that such high precision as provided by TTRG is somewhat pointless, since the horizontally homogeneous turbid medium model (on which is based the LTE) is already very approximate, and so are all our data (leaf and ground optical coefficients, leaf area densities, etc.). However, since TTRG is anyhow faster, one might as well use it. We thereby know at least that integration of the LTE produces very small errors in the absence of sunlight and emissions (about $10^{-12}$ judging from 'realistic' model computations with semi-isotropic leaves), and is of amply sufficient precision when sunlight is present (and treated as an 'emission'), namely better than $10^{-3}$ for all relevant values $0.05 \le \mu_h \le 1$, if we use thin layers of LAI $\ell \approx 0.1$ (the drop in precision being due to the Simpson numerical integration of the propagated emissions).

The case of extreme light trapping provides an example for which iterative integration requires large numbers of iterations, hence large computation times. Moreover, it reveals that specifying a convergence criterion and proper thin layer thickness in the iterative method can sometimes be problematic. Of course, realistic canopies do not suffer these problems. But one should be aware that iteration *can* become impractical in certain cases. Our TTRG method remains as efficient, while also being faster and more accurate in realistic situations.

In this paper we assumed that the diffuse radiances are $\varphi$-symmetric, so that it sufficed to use $\varphi$-symmetric light sectors ($N_J = 18$ sectors, specifically). In that case, iterative integration and TTRG both take little computer time for realistic canopies, so that computation time is not a serious issue here. However, if we wish to compute diffuse radiances which vary with $\varphi$ (as should really be done since photo-synthesis is highly non-linear in the radiances), then one should use 18 azymuthal sectors, say, for a total of $N_J = 18 \times 18 = 324$ light sectors. Then computation time becomes an issue, since it scales like $N_J^2$.

**Appendix A**: **Approximations for exponentials**

We here show that the relative errors done in approximating $e^x \approx (1 \pm x/N)^{\pm N}$ have the following expansions in powers of $N^{-1}$:

$$\begin{aligned}\delta_+(x,N) &\equiv \frac{e^x - (1+x/N)^N}{e^x} = \tfrac{1}{2}x^2N^{-1} - \left(\tfrac{1}{3}x^3 + \tfrac{1}{8}x^4\right)N^{-2} + \dots \\ \delta_-(x,N) &\equiv \frac{(1-x/N)^{-N} - e^x}{e^x} = -\delta_+(x,-N) = \tfrac{1}{2}x^2N^{-1} + \left(\tfrac{1}{3}x^3 + \tfrac{1}{8}x^4\right)N^{-2} + \dots\end{aligned} \tag{A.1}$$

*Proof*: Note first that $\delta_+(x,N) = 1 - A(x)$, where

$$A(x) \equiv e^{-x}(1+x/N)^N = 1 - \tfrac{1}{2}\tfrac{1}{N}x^2 + \tfrac{1}{3}\tfrac{1}{N^2}x^3 + \tfrac{1}{8}\left(\tfrac{1}{N^2} + \tfrac{2}{N^3}\right)x^4 + \dots \tag{A.2}$$

the expansion being obtained by multiplying out the two series $e^{-x} = 1 - x + \frac{1}{2}x^2 - \dots$ and $(1+y)^N = 1 + Ny + \frac{1}{2}N(N-1)y^2 + \dots$, where $y \equiv x/N$. Although (A.1) agrees with (A.2) to the orders shown in $x$ and $N^{-1}$, this does not guarantee that the coefficients of $N^{-1}$, $N^{-2}$,..., in (A.1) do not contain higher powers of $x$. To get (A.1), write $(1+y)^N = (e^y - \varepsilon)^N$ where $\varepsilon = \frac{1}{2!}y^2 + \frac{1}{3!}y^3 + \dots$, and then compute, noting that $e^{Ny} = e^x$:

$$\begin{aligned}A(x) &= e^{-x}(e^y - \varepsilon)^N = e^{-x}\left(e^{Ny} - Ne^{(N-1)y}\varepsilon + \tfrac{1}{2}N(N-1)e^{(N-2)y}\varepsilon^2 - \dots\right) \\ &= 1 - Ne^{-y}\varepsilon + \tfrac{1}{2}N(N-1)e^{-2y}\varepsilon^2 + \dots \qquad (y \equiv x/N)\end{aligned} \tag{A.3}$$

Since $\varepsilon \sim N^{-2}$, we see that the next power in $\varepsilon$ goes as $N^3\varepsilon^3 \sim N^{-3}$, etc. So to order $N^{-2}$:

$$\begin{aligned}A(x) &= 1 - Ne^{-y}(\tfrac{1}{2}y^2 + \tfrac{1}{6}y^3) + \tfrac{1}{2}N(N-1)e^{-2y}\tfrac{1}{4}y^4 + \dots \\ &= 1 - \tfrac{1}{2}e^{-y}\tfrac{x^2}{N} - \tfrac{1}{6}e^{-y}\tfrac{x^3}{N^2} + \tfrac{1}{8}\tfrac{N-1}{N}e^{-2y}\tfrac{x^4}{N^2} + \dots \\ &= 1 - \tfrac{1}{2}(1 - \tfrac{x}{N})\tfrac{x^2}{N} - \tfrac{1}{6}\tfrac{x^3}{N^2} + \tfrac{1}{8}\tfrac{x^4}{N^2} + \dots = 1 - \tfrac{1}{2}\tfrac{x^2}{N} + \tfrac{1}{3}\tfrac{x^3}{N^2} + \tfrac{1}{8}\tfrac{x^4}{N^2} + \dots\end{aligned} \tag{A.4}$$

whence (A.1). For example, with $x = 2$ and $N = 100$, the exact $\delta_+(x,N)$ is $0.019543\ldots \approx 2\%$, while (A.1) yields $\frac{1}{2}x^2/N = \frac{1}{2} \times 4/100 = 0.02$ to order $N^{-1}$, and $0.019533\ldots$ to order $N^{-2}$.

Note that $\left(1+x/N\right)^N < e^x < \left(1-x/N\right)^{-N}$, since $1 \pm x/N \approx e^{\pm x/N} - \frac{1}{2}\left(x/N\right)^2$. For instance, if $x = 10$ and $N = 100$ [$N = 1000$], then $\left(1+x/N\right)^N \approx 13\,780$ [$\approx 20\,959$], and $\left(1-x/N\right)^{-N} \approx 37\,648$ [$\approx 23\,164$], while $e^{10} \approx 22\,026$.

**Appendix B: The functions** $g_\pm(\mu,\mu')$ **and** $g_\pm(\Delta\mu,\mu')$

**B.1** We write the scalar product of two unit vectors $\Omega=(\theta,\varphi)$ and $\Omega'=(\theta',\varphi')$ as

$$\begin{aligned}&\Omega\cdot\Omega'=a+b\cos(\varphi-\varphi'),\qquad a=\mu\mu',\qquad b=\beta\beta'\geq 0\\&\mu=\cos\theta,\qquad \beta\equiv\sqrt{1-\mu^2}=\sin\theta\geq 0\end{aligned}\tag{B.1}$$

Recall that $\theta\in(0,\pi)\Rightarrow\sin\theta\geq 0$. We will use the inverse trigonometric functions

$$\begin{aligned}&\cos^{-1}x\in(0,\pi),\qquad \sin^{-1}x\in(-\pi/2,\pi/2),\qquad \tan^{-1}x\in(-\pi/2,\pi/2)\\&\cos^{-1}(-x)=\pi-\cos^{-1}x,\qquad \sin^{-1}(-x)=-\sin^{-1}x,\qquad \tan^{-1}(-x)=-\tan^{-1}x\end{aligned}\tag{B.2}$$

Using $b^2-a^2=1-\mu'^2-\mu^2$, we note the following derivatives:

$$\begin{aligned}&\frac{d}{d\mu}(b^2-a^2)=-2\mu,\qquad \frac{d}{d\mu}(a/b)=\frac{a/b}{\mu(1-\mu^2)}\\&\frac{d}{d\mu}\cos^{-1}\left(-\frac{a}{b}\right)=\frac{\mu'}{(1-\mu^2)\sqrt{b^2-a^2}},\qquad \frac{d}{d\mu}\tan^{-1}\left(\frac{\mu}{\sqrt{b^2-a^2}}\right)=\frac{1}{\sqrt{b^2-a^2}}\end{aligned}\tag{B.3}$$

**B.2** ***The function*** $G_{\mu'}(\mu)$: As in Eq.(7) in Ref.[2], we define the function

$$\zeta(\mu)=\sqrt{1-\mu^2}-\mu\cos^{-1}\mu=-\int d\mu\cos^{-1}\mu\tag{B.4}$$

We also define $G_{\mu'}(\mu)$ as the indefinite integral of $\frac{1}{\pi}b\zeta(-a/b)$, for fixed $\mu'$, as follows:

$$\dot{G}_{\mu'}(\mu)\equiv\tfrac{1}{\pi}b\zeta(-a/b)=\tfrac{1}{\pi}\sqrt{b^2-a^2}+\tfrac{1}{\pi}a\cos^{-1}(-a/b)=(d/d\mu)G_{\mu'}(\mu)\tag{B.5}$$

$$\begin{aligned}G_{\mu'}(\mu)\equiv\int d\mu\,\dot{G}_{\mu'}(\mu)=&\frac{1}{2\pi}\left[\mu\sqrt{b^2-a^2}-\beta'^2\cos^{-1}\left(\frac{\mu}{\beta'}\right)\right]\\&+\frac{1}{2\pi}\mu'\left[\mu'\tan^{-1}\left(\frac{\mu}{\sqrt{b^2-a^2}}\right)-(1-\mu^2)\cos^{-1}\left(-\frac{a}{b}\right)\right]\end{aligned}\tag{B.6}$$

as is readily verified using (B.3). Using (B.2), we find that

$$\begin{aligned}&G_{-\mu'}(\mu)=G_{\mu'}(\mu)+\tfrac{1}{2}\mu'(1-\mu^2)\\&G_{\mu'}(-\mu)=-G_{\mu'}(\mu)-\tfrac{1}{2}\left[\mu'(1-\mu^2)+(1-\mu'^2)\right]\end{aligned}\tag{B.7}$$

Note the special cases:

$$\begin{aligned}&G_{\mu'\geq 0}(\beta')=\tfrac{1}{4}\mu'^2-\tfrac{1}{2}\mu'^3,\qquad G_{\mu'\geq 0}(-\beta')=\tfrac{1}{4}\mu'^2-\tfrac{1}{2}\qquad &(a)\\&G_0(\mu)=\tfrac{1}{2\pi}\left[\mu\sqrt{1-\mu^2}-\cos^{-1}\mu\right],\qquad G_{\mu'}(0)=\tfrac{1}{4}(\mu'^2-\mu'-1)\qquad &(b)\\&G_0(1)=0,\qquad G_0(-1)=-\tfrac{1}{2},\qquad G_{\pm 1}(0)=\mp\tfrac{1}{4}\qquad &(c)\end{aligned}\tag{B.8}$$

*Proof*: (a) $\mu = \pm\beta'$ and $\mu' \geq 0$ imply $1-\mu^2 = \mu'^2$, $b = \mu'\beta$, $a/b = \pm 1$ and $b^2 - a^2 = 0$, so that $\tan^{-1}\left(\mu/\sqrt{b^2-a^2}\right) = \tan^{-1}(\pm\infty) = \pm\frac{\pi}{2}$, $\cos^{-1}(-a/b) = \cos^{-1}(\mp 1) = \left\{\begin{smallmatrix}\pi\\0\end{smallmatrix}\right\}$, and $\cos^{-1}(\mu/\beta') = \cos^{-1}(\pm 1) = \left\{\begin{smallmatrix}0\\\pi\end{smallmatrix}\right\}$. (*b*) $\mu' = 0$ ($\mu = 0$) implies $a = 0$ and $b = \beta$ ($b = \beta'$). (*c*) follows from (*b*). Note, since $G_{\mu'}(\mu)$ is real only for $|a| \leq b$, that if $\mu = \pm 1$ or $\mu' = \pm 1$, implying $b = 0$, then $G_{\mu'}(\mu)$ is real only for $\mu' = 0$ or $\mu = 0$ (as in (*c*)). Note also that $G_{\mu'\neq 0}(1)$ is complex, while $G_{\mu'}(\mu \neq 0)$ diverges as $\mu' \to 1$.[9]

**B.3** ***The functions*** $g_\pm(\mu,\mu')$: We define, for $\mu \in (-1,1)$, $\mu' \in (-1,1)$:

$$g_\pm(\mu,\mu') = \tfrac{1}{2\pi}\int_{\pm\Omega\cdot\Omega'>0} d\varphi |\Omega\cdot\Omega'| = \tfrac{1}{2\pi}\int_{-\pi}^{\pi} d\varphi \left| a + b\cos\varphi \right|_\pm \tag{B.9}$$

using notation (1). Note the integrals, and special cases:

$$\int_{-1}^{1} d\mu\, g_\pm(\mu,\mu') = \tfrac{1}{2\pi}\int_{\pm\Omega\cdot\Omega'>0} d\Omega |\Omega\cdot\Omega'| = \tfrac{1}{2\pi}\int_{\pm\Omega_z>0} d\Omega |\Omega_z| = \tfrac{1}{2} \tag{B.10}$$

$$\begin{aligned} g_\pm(\mu,0) &= \tfrac{1}{2\pi}\int_{-\pi}^{\pi} d\varphi\, \beta |\cos\varphi|_+ = \tfrac{1}{2\pi}\beta\int_{-\pi/2}^{\pi/2} d\varphi\, \cos\varphi = \beta/\pi \\ g_\pm(\mu,1) &= \tfrac{1}{2\pi}\int_{-\pi}^{\pi} d\varphi\, |\mu|_\pm = |\mu|_\pm \end{aligned} \tag{B.11}$$

Note also the symmetries

$$\begin{aligned} g_\pm(\mu,\mu') &= g_\pm(\mu',\mu) = g_\pm(-\mu,-\mu') \qquad (i) \\ g_-(\mu,\mu') &= g_+(-\mu,\mu') = g_+(\mu,-\mu') \qquad (ii) \end{aligned} \tag{B.12}$$

where (*ii*) follows from $a + b\cos\varphi \to -\left(a + b\cos(\varphi+\pi)\right)$ if $(\mu,\mu') \to (-\mu,\mu')$. Because of (*ii*), it suffices to evaluate $g_+(\mu,\mu')$ for $-1 \leq \mu \leq 1$ and $\mu' \geq 0$. Now, if $a \leq -b$, then $a + b\cos\varphi \leq 0$ for all $\varphi$, hence $g_+ = 0$; if $a \geq b$, then $a + b\cos\varphi \geq 0$ for all $\varphi$, hence $g_+ = \frac{1}{2\pi}\int_{-\pi}^{\pi} d\varphi (a + b\cos\varphi) = a$; finally, if $|a| \leq b$, then $a + b\cos\varphi \geq 0$ for $-\varphi_c \leq \varphi \leq \varphi_c$, where $\varphi_c \equiv \cos^{-1}(-a/b)$, so that $g_+ = \frac{1}{2\pi}\int_{-\varphi_c}^{\varphi_c} d\varphi (a + b\cos\varphi) = \pi^{-1}(a\varphi_c + b\sin\varphi_c)$. Hence, noting that $\sin\varphi_c = b^{-1}\sqrt{b^2 - a^2}$:

---

[9] Considering $\cos^{-1} L = z$, $\cos z = \frac{1}{2}\left(e^{iz} + e^{-iz}\right) = L$, we see that $\left|\cos^{-1} L\right| \sim \log|L|$ as $|L| \to \infty$. It follows that $(1-\mu^2)\cos^{-1}(-a/b)$ vanishes as $\mu \to 1$, and diverges as $\mu' \to 1$.

$$g_+(\mu,\mu')=\begin{cases}0 & if \quad a\le -b\\ \frac{1}{\pi}\sqrt{b^2-a^2}+\frac{1}{\pi}a\cos^{-1}\left(-a/b\right)=\frac{1}{\pi}b\zeta\left(-a/b\right) & if \quad a^2\le b^2\\ a & if \quad a\ge b\end{cases} \tag{B.13}$$

as also follows from (B.4),(B.9) in Ref.[2]. Noting now that $b^2-a^2=\beta'^2-\mu^2$, we see that $a^2\le b^2 \Leftrightarrow -\beta'\le\mu\le\beta'$; also, $a^2\ge b^2\Leftrightarrow \hat{\mu}\ge\beta'$, so that if $\mu'\ge 0$, then $a\ge b\Leftrightarrow \mu\ge\beta'$, and $a\le -b\Leftrightarrow \mu\le-\beta'$. It follows that (B.13), for $\mu'\ge 0$, may be rewritten as, using (B.5):

$$g_+(\mu'\ge 0,\mu)=\begin{cases}0 & if \quad \mu\le-\beta'\\ \dot{G}_{\mu'}(\mu) & if \quad -\beta'\le\mu\le\beta'\\ \mu\,\mu' & if \quad \mu\ge\beta'\end{cases} \qquad \left(\mu'\ge 0\right) \tag{B.14}$$

**A.4** ***The functions*** $g_\pm(\Delta\mu,\mu')$: We next evaluate, for $\Delta\mu=(\mu_1,\mu_2)\subset(-1,1)$, $\mu_1\le\mu_2$:

$$g_\pm(\Delta\mu,\mu')\equiv\int_{\mu_1}^{\mu_2}d\mu\, g_\pm(\mu,\mu')=\tfrac{1}{\pi}\int_{\mu_1}^{\mu_2}d\mu\int_0^\pi d\varphi\left|a+b\cos\varphi\right|_\pm \tag{B.15}$$

If the intervals $\Delta\mu$ partition $(-1,1)$, then by (B.10):

$$\sum\nolimits_{\Delta\mu} g_\pm(\Delta\mu,\mu')=\int_{-1}^{1}d\mu\, g_\pm(\mu,\mu')=\tfrac{1}{2} \tag{B.16}$$

Since $g_+(\mu,\mu')=g_-(\mu,-\mu')=g_+(-\mu,-\mu')$ by (B.12), we have the symmetries

$$g_+(\Delta\mu,-\mu')=g_-(\Delta\mu,\mu')=g_+(-\Delta\mu,\mu'), \qquad -\Delta\mu\equiv(-\mu_2,-\mu_1) \tag{B.17}$$

where we denote $-\Delta\mu\equiv(-\mu_2,-\mu_1)$. It thus suffices to evaluate $g_+(\Delta\mu,\mu')$ for $\mu'\ge 0$. Using then (B.14), we have for instance, for $\mu_1\le-\beta'\le\beta'\le\mu_2$:

$$\begin{aligned}g_+(\Delta\mu,\mu')&=\int_{\mu_1}^{-\beta'}d\mu g_+ +\int_{-\beta'}^{\beta'}d\mu g_+ +\int_{\beta'}^{\mu_2}d\mu g_+ =0+\int_{-\beta'}^{\beta'}d\mu\,\dot{G}(\mu)+\mu'\int_{\beta'}^{\mu_2}d\mu\,\mu\\ &=\left[G(\beta')-G(-\beta')\right]+\tfrac{1}{2}\mu'\left(\mu_2^2-\beta'^2\right)\end{aligned} \tag{B.18}$$

We thereby find, for $\mu_1\le\mu_2$ and $\mu'\ge 0$:

$$g_+(\Delta\mu,\mu')=\begin{cases}0 & if \quad \mu_2\le-\beta' & (a)\\ \frac{1}{2}\mu'(\mu_2^2-\mu_1^2) & if \quad \mu_1\ge\beta' & (b)\\ G_{\mu'}(\beta')-G_{\mu'}(-\beta')+\frac{1}{2}\mu'(\mu_2^2-\beta'^2) & if \quad \mu_1\le-\beta'\le\beta'\le\mu_2 & (c)\\ G_{\mu'}(\mu_2)-G_{\mu'}(-\beta') & if \quad \mu_1\le-\beta'\le\mu_2\le\beta' & (d)\\ G_{\mu'}(\beta')-G_{\mu'}(\mu_1)+\frac{1}{2}\mu'(\mu_2^2-\beta'^2) & if \quad -\beta'\le\mu_1\le\beta'\le\mu_2 & (e)\\ G_{\mu'}(\mu_2)-G_{\mu'}(\mu_1) & if \quad -\beta'\le\mu_1\le\mu_2\le\beta' & (f)\end{cases} \tag{B.19}$$

Note that (B.19) excludes the problematic values $G_1(\mu \neq 0)$ and $G_{\mu'\neq 0}(\pm 1)$ (as should be since (B.15) is well defined). With $G_{\mu'}(\pm\beta')$ given by (B.8)(*a*), case (*c*) also reads $g_+(\Delta\mu,\mu') = \frac{1}{2} - \frac{1}{2}\mu'(1-\mu_2^2)$, whence, for $\Delta\mu = (-1,1)$, $g_+\big((-1,1),\mu'\big) = \frac{1}{2} = \int_{-1}^{1} d\mu\, g_+(\mu,\mu')$, as should be by (B.10). We also find, for $0 \le \mu' \le 1$, that $g_+\big(\pm(0,1),\mu'\big) = \frac{1}{4}(1 \pm \mu')$, where $-(0,1) \equiv (-1,0)$, hence, by (B.17), for any $\mu'$:

$$g_\pm\big((0,1),\mu'\big) = \int_0^1 d\mu\, g_\pm(\mu,\mu') = \tfrac{1}{4}(1 \pm \mu') \qquad (-1 \le \mu' \le 1) \tag{B.20}$$

Whence again $g_\pm\big((-1,1),\mu'\big) = g_\pm\big((-1,0),\mu'\big) + g_\pm\big((0,1),\mu'\big) = \frac{1}{2}$.

**Appendix C: Simpson's integration rule**

Consider the integral (for any real $a$ and $h$):

$$I = \int_{a-h}^{a+h} dz\, f(z) = h\int_{-1}^{1} dx\, y(x), \qquad z \equiv a + hx, \qquad y(x) \equiv f(a+hx) \tag{C.1}$$

Denote

$$y_0 \equiv y(0) = f(a), \qquad y_\pm \equiv y(\pm 1) = f(a \pm h), \qquad y_m = \tfrac{1}{2}(y_- + y_+) \tag{C.2}$$

The integral $I$ is approximated by $I \approx I_m = y_m \cdot 2h$ in the 'trapezoid' rule, and by $I \approx I_0 = y_0 \cdot 2h$ in the 'midpoint' rule. In Simpson's rule, $y(x)$ is first approximated by a quadratic polynomial passing through the three points $(-1, y_-)$, $(0, y_0)$ and $(1, y_+)$:

$$y(x) \approx y_S(x) = a + bx + cx^2 = y_0 + \tfrac{1}{2}(y_+ - y_-)x + (y_m - y_0)x^2 \tag{C.3}$$

Since the linear term $bx$ contribute zero to the integral over $(-1,1)$, we get

$$I \approx I_S \equiv h\int_{-1}^{1} dx\, y_S(x) = h\int_{-1}^{1} dx\,(y_0 + cx^2) = \tfrac{1}{3}h(y_- + 4y_0 + y_+) \tag{C.4}$$

Simpson's rule is obviously exact if *f*(*x*) is a quadratic polynomial, and also if *f*(*x*) is a cubic polynomial. Indeed, if $y(x) = a_0 + a_1x + a_2x^2 + \ldots$, then $y_0 = a_0$ and $y_\pm = a_0 \pm a_1 + a_2 \pm a_3 \pm \ldots$, so that $y_m = a_0 + a_2 + a_4 + \ldots$ Thus, if $f(x)$ is cubic, then $y_m - y_0 = a_2$, so that $y(x)$ and $y_S(x)$ have identical coefficients for $x^0$ and $x^2$, hence equal integrals over $(-1,1)$ (since the $x^3$ term integrates to $0$).

In our problem we also require the separate integrals over $(-h,0)$ and $(0,h)$:

$$I_\pm \approx I_{S\pm} = \pm\int_0^{\pm h} dz\, f_S(z) = \tfrac{1}{2}I_S \pm h\int_0^{\pm 1} dx\, bx = \tfrac{1}{12}h(5y_\pm + 8y_0 - y_\mp) \tag{C.5}$$

(of course, $I_{S+} + I_{S-} = I_S$). These 'half integrals' are again exact if $f(x)$ is a quadratic, but no longer if $f(x)$ is a cubic. Now, by the 'interpolation error theorem' (see, e.g., Ref.[7]), we have:

$$\left| y_S(x) - y(x) \right| \le \tfrac{1}{4!} \left| y'''\big(p(x)\big)(x+1)x(x-1) \right| \le \tfrac{1}{4!} M_3 h^3 \left| x^3 - x \right| \qquad \text{(C.6)}$$

for some $p(x) \in (-1,1)$, and $M_k$ is a bound on the $k$th derivative of $f(z)$ inside $(-h,h)$. Thus the error for the half-integrals (C.5) has the bound

$$Error \le \tfrac{1}{4!} M_3 h^4 \int_0^1 dx\,(x - x^3) = \tfrac{1}{96} M_3 h^4 \qquad \text{(C.7)}$$

This is in general larger than the error bound $\frac{1}{90} M_4 h^5$ for the integral (C.4) over $(-h,h)$. We use (C.4) and (C.5) alternately to integrate the 'propagated emissions' (89) up to even and odd numbered thin layers, respectively, whence an alternance in the precision. For instance, with $h = \ell = 0.1$, and $\mu_h \ge 0.3$, we find that the computed radiances just after odd-numbered thin layers are roughly 1 digit less accurate than radiances just after even-numbered thin layers.